\documentclass[twocolumn]{aastex631}
\pdfoutput=1
\definecolor{darkblue}{rgb}{0.0,0.0,0.4}
\definecolor{darkgreen}{rgb}{0.0,0.4,0.0}
\hypersetup{linkcolor=black, citecolor=darkgreen, urlcolor=darkblue}

\received{May 16, 2023}
\revised{August 11, 2023}
\accepted{August 28, 2023}

\submitjournal{ApJ}

\usepackage{savesym}
\savesymbol{tablenum}
\usepackage{siunitx}
\restoresymbol{SIX}{tablenum}

\DeclareSIUnit \Msun {M_{\odot}}
\DeclareSIUnit \Lsun {L_{\odot}}
\DeclareSIUnit \LsunV {L_{\odot V}}
\DeclareSIUnit \pc {pc}
\DeclareSIUnit \kpc {kpc}
\DeclareSIUnit \mas {mas}
\DeclareSIUnit \yr {yr}
\DeclareSIUnit \pixel {pix}
\DeclareSIUnit \SN {SN}
\DeclareSIUnit \mag{mag}

\shorttitle{oMEGACat I}
\shortauthors{Nitschai et al.}

\begin{document}

\title{oMEGACat I: MUSE spectroscopy of 300,000 stars within the half-light radius of $\omega$~Centauri}

\correspondingauthor{M. S. Nitschai}
\email{nitschai@mpia.de}

\author[0000-0002-2941-4480]{M. S. Nitschai}
\affiliation{Max Planck Institute for Astronomy, K\"onigstuhl 17, D-69117 Heidelberg, Germany}
\author[0000-0002-6922-2598]{N. Neumayer}
\affiliation{Max Planck Institute for Astronomy, K\"onigstuhl 17, D-69117 Heidelberg, Germany}
\author[0009-0005-8057-0031]{C. Clontz}
\affiliation{Max Planck Institute for Astronomy, K\"onigstuhl 17, D-69117 Heidelberg, Germany}
\affiliation{Department of Physics and Astronomy, University of Utah, Salt Lake City, UT 84112, USA}
\author[0000-0002-5844-4443]{M. H\"aberle}
\affiliation{Max Planck Institute for Astronomy, K\"onigstuhl 17, D-69117 Heidelberg, Germany}
\author[0000-0003-0248-5470]{A. C. Seth}
\affiliation{Department of Physics and Astronomy, University of Utah, Salt Lake City, UT 84112, USA}
\author[0000-0003-2466-5077]{T.-O. Husser}
\affiliation{Institut für Astrophysik und Geophysik, Georg-August-Universität Göttingen, Friedrich-Hund-Platz 1, 37077 Göttingen, Germany}
\author[0000-0001-6604-0505]{S. Kamann}
\affiliation{Astrophysics Research Institute, Liverpool John Moores University, 146 Brownlow Hill, Liverpool L3 5RF, UK}
\author[0000-0002-1212-2844]{M. Alfaro-Cuello}
\affiliation{Facultad de Ingenier\'{i}a y Arquitectura, Universidad Central de Chile, Av. Francisco de Aguirre 0405, La Serena, Coquimbo, Chile}
\affiliation{Space Telescope Science Institute, 3700 San Martin Drive, Baltimore, MD 21218, USA}
\author{N. Kacharov}
\affiliation{Leibniz Institute for Astrophysics, An der Sternwarte 16, 14482 Potsdam, Germany}
\author[0000-0003-3858-637X]{A. Bellini}
\affiliation{Space Telescope Science Institute, 3700 San Martin Drive, Baltimore, MD 21218, USA}
\author{A. Dotter}
\affiliation{Department of Physics and Astronomy, Dartmouth College, Hanover, NH 03755 US}
\author[0000-0001-6187-5941]{S. Dreizler}
\affiliation{Institut für Astrophysik und Geophysik, Georg-August-Universität Göttingen, Friedrich-Hund-Platz 1, 37077 Göttingen, Germany}
\author[0000-0002-0160-7221]{A. Feldmeier-Krause}
\affiliation{Max Planck Institute for Astronomy, K\"onigstuhl 17, D-69117 Heidelberg, Germany}
\author[0000-0002-7547-6180]{M. Latour}
\affiliation{Institut für Astrophysik und Geophysik, Georg-August-Universität Göttingen, Friedrich-Hund-Platz 1, 37077 Göttingen, Germany}
\author[0000-0001-9673-7397]{M. Libralato}
\affiliation{AURA for the European Space Agency (ESA), Space Telescope Science Institute, 3700 San Martin Drive, Baltimore, MD 21218, USA}
\author[0000-0001-7506-930X]{A. P. Milone}
\affiliation{Dipartimento di Fisica e Astronomia “Galileo Galilei,” Univ. di Padova, Vicolo dell’Osservatorio 3, Padova, I-35122, Italy}
\author[0000-0002-1670-0808]{R. Pechetti}
\affiliation{Astrophysics Research Institute, Liverpool John Moores University, 146 Brownlow Hill, Liverpool L3 5RF, UK}
\author[0000-0003-4546-7731]{G. van de Ven}
\affiliation{Department of Astrophysics, University of Vienna, T\"urkenschanzstrasse 17, 1180 Wien, Austria}
\author[0000-0001-6215-0950]{K. Voggel}
\affiliation{Universit\'{e} de Strasbourg, CNRS, Observatoire astronomique de Strasbourg, UMR 7550, F-67000 Strasbourg, France}
\author[0000-0002-6442-6030]{Daniel R. Weisz}
\affiliation{Department of Astronomy, University of California, Berkeley, CA, 94720, USA}

\begin{abstract}

Omega Centauri ($\omega$~Cen) is the most massive globular cluster of the Milky Way and has been the focus of many studies that reveal the complexity of its stellar populations and kinematics. However, most previous studies have used photometric and spectroscopic datasets with limited spatial or magnitude coverage, while we aim to investigate it having full spatial coverage out to its half-light radius and stars ranging from the main sequence to the tip of the red giant branch. This is the first paper in a new survey of $\omega$~Cen that combines uniform imaging and spectroscopic data out to its half-light radius to study its stellar populations, kinematics, and formation history. In this paper, we present an unprecedented MUSE spectroscopic dataset combining 87 new MUSE pointings with previous observations collected from guaranteed time observations. We extract spectra of more than 300,000 stars reaching more than two magnitudes below the main sequence turn-off. We use these spectra to derive metallicity and line-of-sight velocity measurements and determine robust uncertainties on these quantities using repeat measurements. Applying quality cuts we achieve signal-to-noise ratios of 16.47/73.51 and mean metallicity errors of 0.174/0.031~dex for the main sequence stars (\SI{18}{\mag} $\rm < mag_{F625W}<$\SI{22}{\mag}) and red giant branch stars (\SI{16}{\mag} $<\rm  mag_{F625W}<$\SI{10}{\mag}), respectively. We correct the metallicities for atomic diffusion and identify foreground stars. This massive spectroscopic dataset will enable future studies that will transform our understanding of $\omega$~Cen, allowing us to investigate the stellar populations, ages, and kinematics in great detail.

\end{abstract}

\keywords{Galaxy nuclei(609) --- Globular star clusters(656) --- Star clusters(1567)}

\section{Introduction} \label{sec:intro}

\begin{figure*}[t!]
\plotone{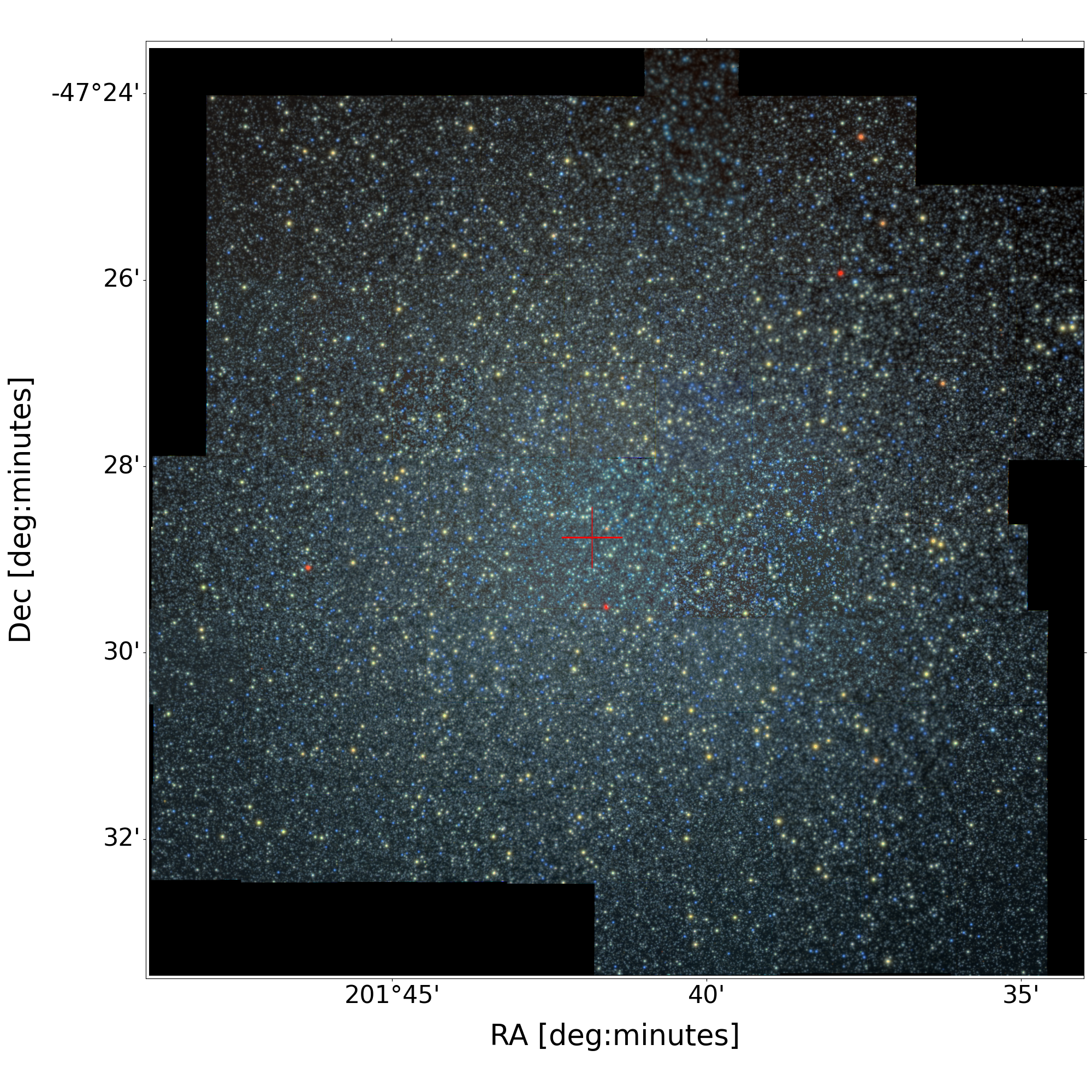}
\caption{\textbf{Image of $\omega$~Cen.} A three-color RGB image of $\omega$~Cen created from MUSE WFM data using synthetic SDSS i, r, and g filters. The image displays the coverage of all WFM data (both GTO and GO). The red cross indicates the center of the cluster.  \label{fig:color im}}
\end{figure*}

In recent years, we have been able to start decoding the formation history of our Galaxy, thanks to large spectroscopic surveys, like the \textit{Gaia} mission \citep{Gaia_mission_16}, APOGEE \citep{Allende_2008, Majewski2017}, LAMOST/LEGUE \citep{Lamost2012} and GALAH \citep{DeSilva_2015, Buder2021}. These surveys, combined with the framework established through cosmological hydrodynamic simulations, have revealed the role that mergers have played in the Milky Way's formation history. 

When less massive satellite galaxies merge with a larger host galaxy, several things happen: The tidal effects of the host galaxy will cause the smaller galaxy to disrupt and form stellar streams \citep[]{Helmi_2001, Mayer_2002}. In addition, dynamical friction can cause the denser central regions of the smaller galaxy to sink toward the center of the host galaxy before being fully disrupted.

The Milky Way has had a quiet merger history, experiencing no recent major mergers \citep{Stewart2008}. The last significant merger happened around 10 Gyr ago with the Gaia-Enceladus satellite \citep[e.g.][]{Haywood2018, Helmi2018}.

Most galaxies, including the Milky Way, contain a dense nuclear star cluster (NSC) at their center \citep{Neumayer_2020}, the densest stellar systems known in the Universe. During a merger event, the NSC of the satellite galaxy will survive without being disrupted, and live on in the halo of the host galaxy \citep[e.g.,][]{Pfeffer_2013}.  

NSCs in low-mass galaxies have very similar properties to massive globular clusters \citep{Georgiev_2014, Neumayer_2020}, with half-light radii of \SIrange{1}{10}{\pc} and masses of \SIrange{e6}{e8}{\Msun}. Without their surrounding host galaxy, these stripped NSCs will look like massive globular clusters. Hence, they will be able to hide amongst the $\gtrsim$ 100 globular clusters in the galaxy. Recent semi-analytic models predict 2--6 stripped nuclei in the halo of the Milky Way \citep{Pfeffer_2014, Kruijssen_2019}. Because masses and star formation histories of NSCs track the galaxies they live in \citep[e.g.][]{Kacharov_2018, Sanchez-Janssen_2019,Fahrion_2021}, stripped nuclei can tell us about the masses of their original hosts and the times of merging if we can identify them. 

An ongoing example of the galaxy merging process is the Sagittarius dwarf spheroidal galaxy, whose stars are wrapped in a stream around the Milky Way, created due to tidal stripping over the last several \si{\giga\yr} \citep{Ibata_1997, Laporte2018}. The nucleus of this galaxy was discovered to be the globular cluster M54 long before the remainder of the galaxy was found. Detailed studies of M54 have revealed that, unlike a typical globular cluster (but typical of NSCs), it has stellar populations with a wide range of metallicities and ages \citep[e.g.][]{Siegel_2007,Alfaro-Cuello_2019}.  

\begin{figure*}[t!]
\plotone{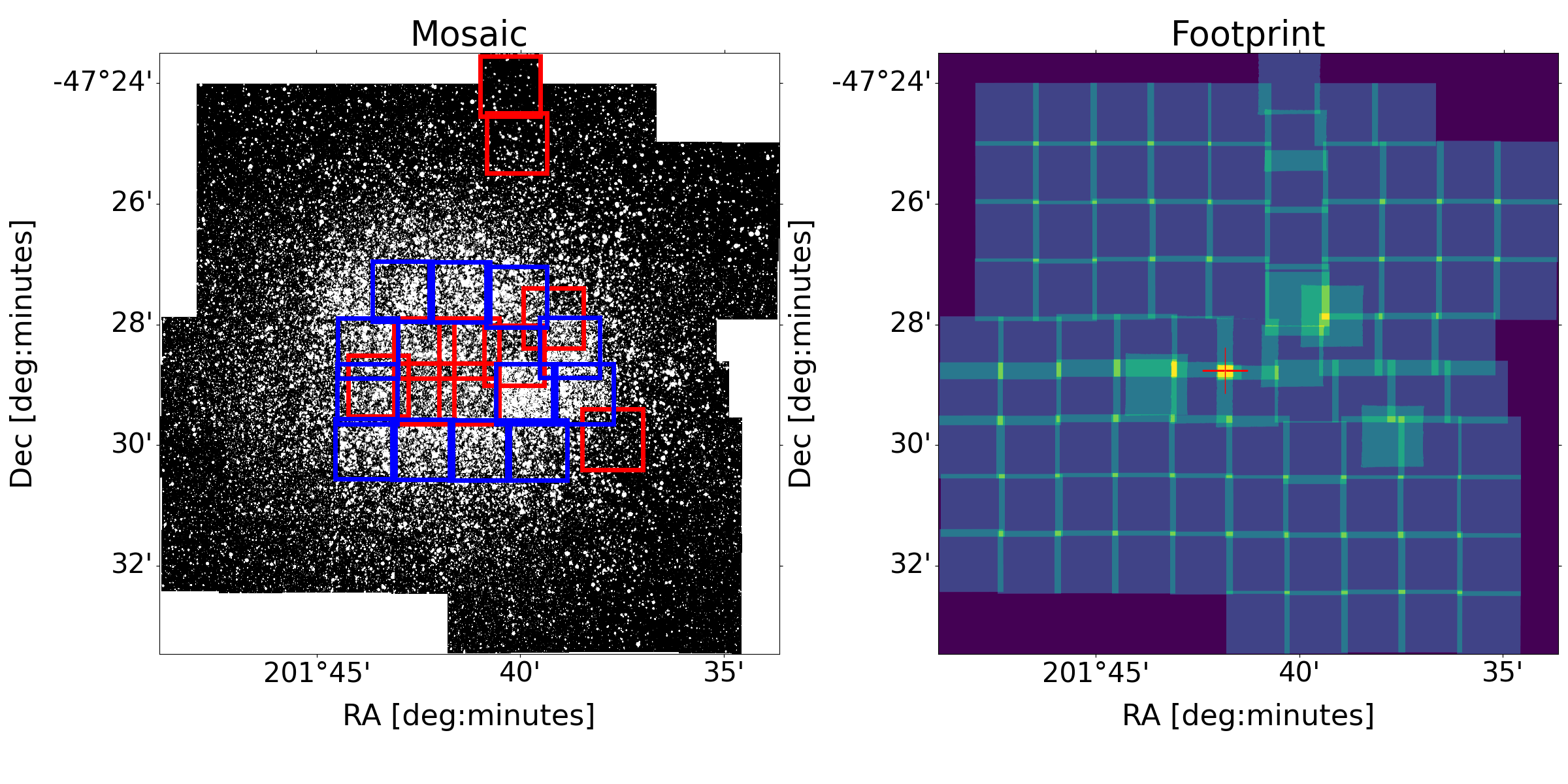}
\caption{\textbf{Image of $\omega$~Cen and footprint of the MUSE pointings.} On the left is a grayscale image created from the combined GO and GTO MUSE WFM data. Overlaid blue colored squares indicate the GO AO pointings, red squares the WFM GTO data, while the region without squares is where the non-AO GO WFM pointings lie. On the right is the footprint of the individual pointings showing the overlap of the data. At the center of the right-hand Figure, the NFM data are within the yellow region at the center of the image, indicated with a red cross.  
\label{fig:data}}
\end{figure*}

$\omega$~Centauri ($\omega$~Cen, NGC 5139) is the brightest, most massive globular cluster of the Milky Way \citep[$\sim$ \SI{3.55e6}{\Msun,}][]{Baumgardt2018}, sitting at a distance of only \SI{5.43}{\kpc} \citep{Baumgardt2021}. Due to the complexity of its stellar populations, it has long been suspected to be the stripped former nucleus of a galaxy that merged with the Milky Way a long time ago \citep[e.g.][]{Bekki_2003}. Evidence for the multiple populations of stars in the cluster comes from the multiple sequences in its color-magnitude diagram \citep[CMD, e.g.][]{Anderson1997, Bedin2004, Bellini_2010, Milone_2017}, a large spread in metallicity \citep[e.g.][]{Freeman1975, Johnson_2010}, and possibly also in age \citep{Hilker_2004, Villanova_2014}. How prolonged the star formation in $\omega$~Cen was is still controversial, i.e., \SIrange{4}{5}{\giga\yr} \citet{Villanova_2007} vs. \SIrange{1}{2}{\giga\yr} \citet{Joo_2013}. Additional evidence for the stripping scenario comes from kinematic information that finds the presence of a central stellar disk and a bias toward tangential orbits in the outer parts \citep{VanDeVen_2006}. $\omega$~Cen's status as stripped nucleus has been further strengthened over the last decade through its association with stellar streams and the Gaia-Enceladus merger \citep[e.g.][]{majewski_2012, Ibata_2019, Limberg2022}. $\omega$~Cen is, therefore, our closest NSC, closer than our own Galactic center, and has much less extinction due to gas and dust. Due to its proximity, it has a large angular size, making it easily observable. For these reasons, $\omega$~Cen is the perfect laboratory for studying NSCs, stripped nuclei and multiple populations. It can also inform us about the merger history of our galaxy since it is likely that its progenitor was part of one of the most massive early mergers of the Milky Way. 

In order to study the cluster in detail and specifically the multiple populations, spectroscopic data are important for metallicity and abundance measurements. Due to the high crowding in the cluster many spectroscopic surveys, like \textit{Gaia} and APOGEE \citep{Meszaros_2021}, are limited to a few bright stars often only at the outskirts of dense clusters like $\omega$~Cen. Other studies like \citet{Johnson_2010, Johnson_2020} have spectroscopic data covering the central region but only for the brightest stars. The largest spectroscopic survey for multiple Galactic globular clusters, including $\omega$~Cen, has been presented in \citet{Kamann_2018}. This survey is covering the very center of the cluster having thousands of spectra from the main sequence (MS) up to the red giant branch (RGB). We aim to further extend that sample out to the half-light radius.

The Multi-Unit Spectroscopic Explorer (MUSE) is the perfect instrument to investigate individual stars in clusters as has been shown in other studies, e.g. \citet{Husser_2016, Kamann_2018, Alfaro-Cuello_2019, Alfaro-Cuello_2020, Kacharov2022}. In this work, we present a large MUSE dataset, covering $\omega$~Cen out to its half-light radius, and the spectroscopic analysis of hundreds of thousands of individual stars, by far the largest spectroscopic dataset ever assembled for $\omega$~Cen, or in any cluster. The paper is divided into the following sections: In \autoref{sec:data} we present the observations and in \autoref{sec:methods} we describe the methods used to get the physical parameters for our catalog. In addition, in \autoref{sec:results} we explain the analysis performed on that catalog to clean and test it. Further tests are also shown in the Appendix. Finally, in \autoref{sec: concl} we summarize and give an outlook on further works in progress using this catalog.

\section{Data} \label{sec:data}

\subsection{Observations and Data Sets} \label{sec:data set}

The data presented in this paper were acquired with MUSE \citep{Bacon2010, Bacon2014}, a second-generation Very Large Telescope (VLT) instrument located at the UT4 at the Paranal Observatory in Chile. Two sets of data are combined in this paper; the first are existing data, part of the MUSE guaranteed time observations (hereafter ``GTO data"), with program IDs: 094.D-0142, 095.D-0629, 096.D-0175, 097.D-0295, 098.D-0148, 099.D-0019, 0100.D-0161, 0101.D-0268, 0102.D-0270, 0103.D-0204, 0104.D-0257, 105.20CR, and 109.23DV. These GTO data consist of 10 pointings with multi-epoch data that have been analyzed already in several papers \citep{Kamann_2018,Husser_2020,Latour_2021} as well as six central pointings using the MUSE narrow field adaptive optics mode (NFM), presented in \citet{Pechetti_2023}. The second dataset is from General Observer (GO) program 105.20CG.001 (PI: N. Neumayer); this includes 87 new MUSE pointings taken between February 2021 and September 2022 -- we refer to this dataset as the ``GO data" and describe this dataset in more detail below. 

MUSE is an integral-field spectrograph, based on image-slicing with 24 identical integral field units (IFUs). The field of view in the wide field mode (WFM), which was used for the GO data, is \SI{59.9}{\arcsecond} $\times$ \SI{60.0}{\arcsecond} for each pointing with a spatial pixel scale of \SI{0.2}{\arcsecond\per\pixel}. The instrument observes in the optical domain (\SI{480}{\nano\meter}- \SI{930}{\nano\meter}) with a resolving power increasing with wavelength from 1770 to 3590 and a spectral sampling of \SI{0.125}{\nano\meter\per\pixel}. For the GTO NFM data, the field of view of each pointing is \SI{7.5}{\arcsecond} $\times$ \SI{7.5} {\arcsecond}, they have a sampling of \SI{0.025}{\arcsecond\per\pixel} and the resolving power is increasing with wavelength from 1740 to 3450. In both the GO and GTO data, some central pointings are also taken using the VLT Adaptive Optics Facility (AOF) \citep{Arsenault2010, Stroebele2012}. In adaptive optics (AO) mode the NaD lines are blocked using a filter, causing a gap in the spectra between \SI{580}{\nano\meter} and \SI{597}{\nano\meter} \citep{MUSE_pipeline}.

In the GO data, each field has three exposures with a rotation of \SI{90}{\degree} between them (no dithering), and the exposure time was \SI{200}{\second}. The observations were requested in service mode at an airmass less than 1.4 and a seeing better than 0$\farcs$8. These conditions were almost always fulfilled, see \autoref{sec:cond data} for more details on the observing conditions. We note that one observing block (OB 1.3) was initially observed using AO with the wrong offset, causing a small gap in our dataset. Therefore, it was repeated without AO in September 2022 once more, to fill in the gap.

We show the spatial coverage of our combined dataset in \autoref{fig:color im} and \autoref{fig:data}. The combined dataset includes a total of 97 WFM pointings as well as six central NFM pointings, providing complete coverage out to the half-light radius of $\omega$~Cen \citep[4.65' or \SI{7.04}{\pc},][]{Baumgardt2018}.   

The reduction and analysis of the GTO data were already done, details on the GTO WFM data are presented in \citet{Kamann_2018}, while the NFM data are presented in \citet{Pechetti_2023}. For the GO data, we duplicated the GTO WFM reduction procedure as closely as possible.  

For the rest of the paper, most of the data reduction and analysis steps refer to just the GO data, except where clearly specified to also include the GTO data.

\subsection{Data Reduction}\label{sec:data red}

We use the MUSE pipeline, version 2.8.3 \citep{MUSE_pipeline,Weilbacher2020}, to reduce the GO data. This pipeline uses CPL \citep{CPL_pipeline} and the \textsc{EsoRex} \citep{ESORex_pipeline} packages and contains all necessary procedures for the data reduction process, including bias subtraction, flat fielding, illumination correction, and wavelength calibration. Further, the pipeline flux calibrates the exposures using a standard star, corrects for the barycentric motion of the Earth, and combines the three single exposures into one for each field. We use the default settings for most parts of the pipeline, except for sky subtraction, where we use a sky continuum file containing zeroes as the model, so that only the emission lines are subtracted, but not the sky continuum. Removing the sky continuum would also remove starlight in a crowded region like ours since even the darkest spaxels still contain starlight. This will leave sky continuum in our final data cubes, but since we do a point spread function-based extraction (see \autoref{sec:PampelMuse}) these emissions will not be in the spectra we extract but in the background components determined during the extraction.

In addition, we are not doing any Raman correction since the field of view is too crowded or the Telluric correction, since our spectral fitting routine can remove them better. Hence, the Telluric lines are included in our spectral fit and are removed then from the spectra, see \autoref{sec:spexxy}.

The total MUSE data coverage is shown in \autoref{fig:data}, where the red squares indicate where the WFM GTO pointings lie and the blue squares where the 12 new WFM-AO pointings are, while all other pointings are WFM without AO, 75 in total. The GTO NFM data are within the yellow region at the center of the right panel of \autoref{fig:data} (the small yellow square where the red cross lies). The gaps in our GO pointings, visible in \autoref{fig:extracted}, in the center and upper part of the image are filled in with the GTO pointings so that we have full coverage of the cluster without gaps when combining the two datasets.

\section{Analysis Methods}\label{sec:methods}

After the reduction of the MUSE mosaic, the next step is to get spectroscopic information for individual stars. Therefore, we need to extract the spectra of individual stars and measure physical parameters from these spectra.

\subsection{Spectral Extraction}\label{sec:PampelMuse}

\begin{figure}[t]
\plotone{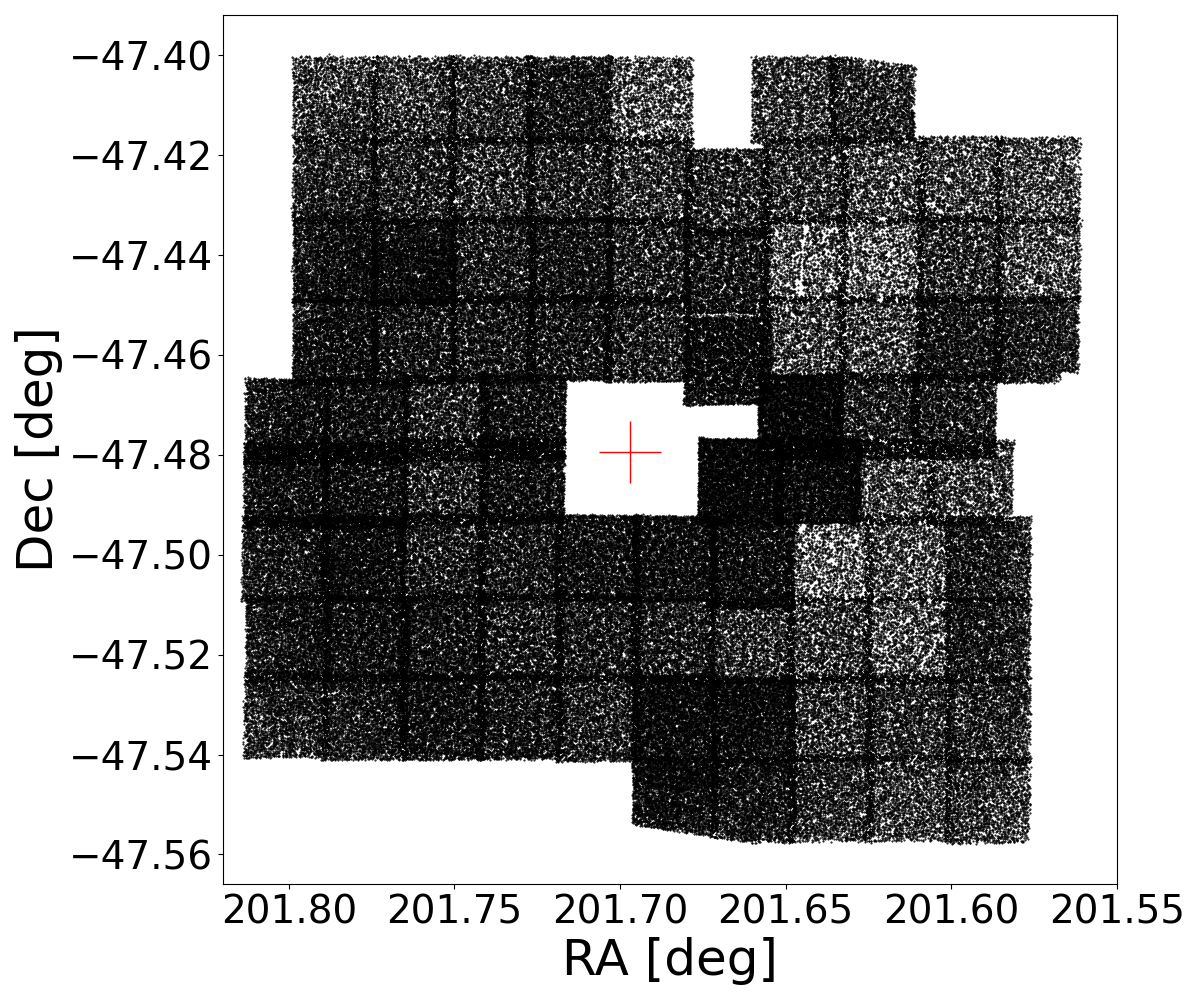}
\caption{\textbf{All extracted GO stars.} The figure shows the positions of all extracted GO stars using the \citet{Anderson2010} photometric catalog. The red cross indicates the center, while the holes at the top and at the center are later filled with the GTO stars. Some edges are cut off because the \textit{HST} footprint does not perfectly cover the MUSE footprint. The overlap regions are denser since more stars are observed due to the fact that there are two or more exposures. The variation in the number of stars seen in different fields is due to varying observing conditions that impact our completeness. \label{fig:extracted}}
\end{figure}

To extract individual spectra for the stars in the MUSE fields we use \textsc{PampelMuse}\footnote{\url{https://PampelMuse.readthedocs.io/en/latest/about.html}} \citep{Kamann2013} and the \textit{Hubble Space Telescope} (\textit{HST}) catalog from \citet{Anderson2010}. \textsc{PampelMuse} takes the photometry and position of the stars in the \textit{HST} catalog as a reference for the stars in the field of view of the MUSE data. It fits the point spread function (PSF) as a function of wavelength. Forced PSF photometry performed at each wavelength for all the stars in the catalog allows \textsc{PampelMuse} to separate sources efficiently even in crowded regions. The \citet{Anderson2010} catalog includes F625W and F435W magnitudes for \num{1.2e6} stars and covers a field of \SI{10}{\arcminute} x \SI{10}{\arcminute} with a pixel scale of \SI{50}{\mas\per\pixel}. We use a Moffat profile \citep{Moffat1969} for the PSF as defined in Section 4.2 of \citet{Kamann2013} and allow it to vary its ellipticity, positional angle, $\beta$-parameter, and full width at half maximum (FWHM), while the initial value for the FWHM is set to be the mean value of the seeing at the time of the observation. The extracted spectra also get a quality flag assigned from \textsc{PampelMuse} where 0 is the best flag a spectrum can get. While there are 4 other criteria for the flags there can also be a combination of them, 1: more than one source contributing, 2: signal-to-noise ratio (SNR) $<$ 10, 4: the flux is negative, and 8: the centroid of the source is outside the field of view of the data cube).

We have performed an additional analysis on some OBs using a different extraction setup and find that the final results for spectra with a SNR $>10$ do not change significantly, see \autoref{sec:PampelMuse test}. In the next section, \autoref{sec:com}, we show the completeness of this extraction for the GO dataset.

In total we extract with \textsc{PampelMuse} 355,682 stars from the GO MUSE dataset, see \autoref{fig:extracted}, where we show their positions.

\subsection{Completeness}\label{sec:com}

\begin{figure*}[t]
\plottwo{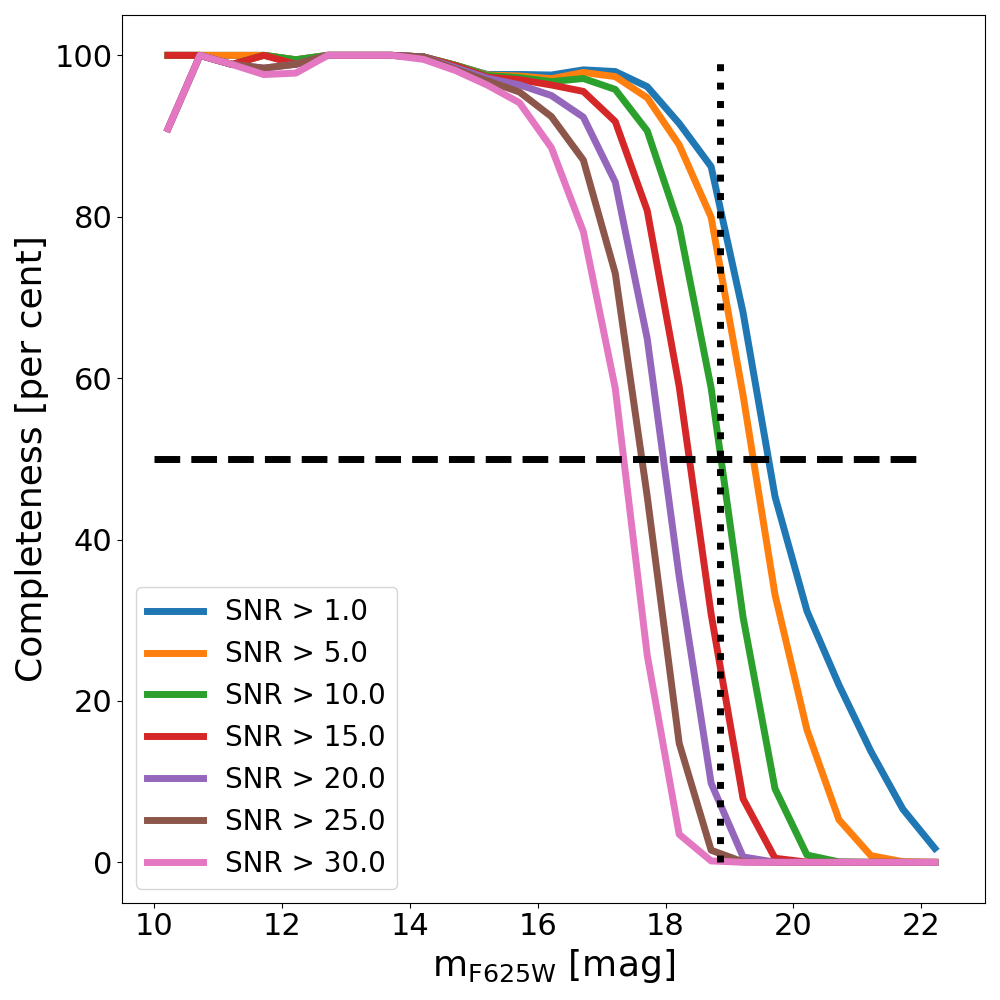}{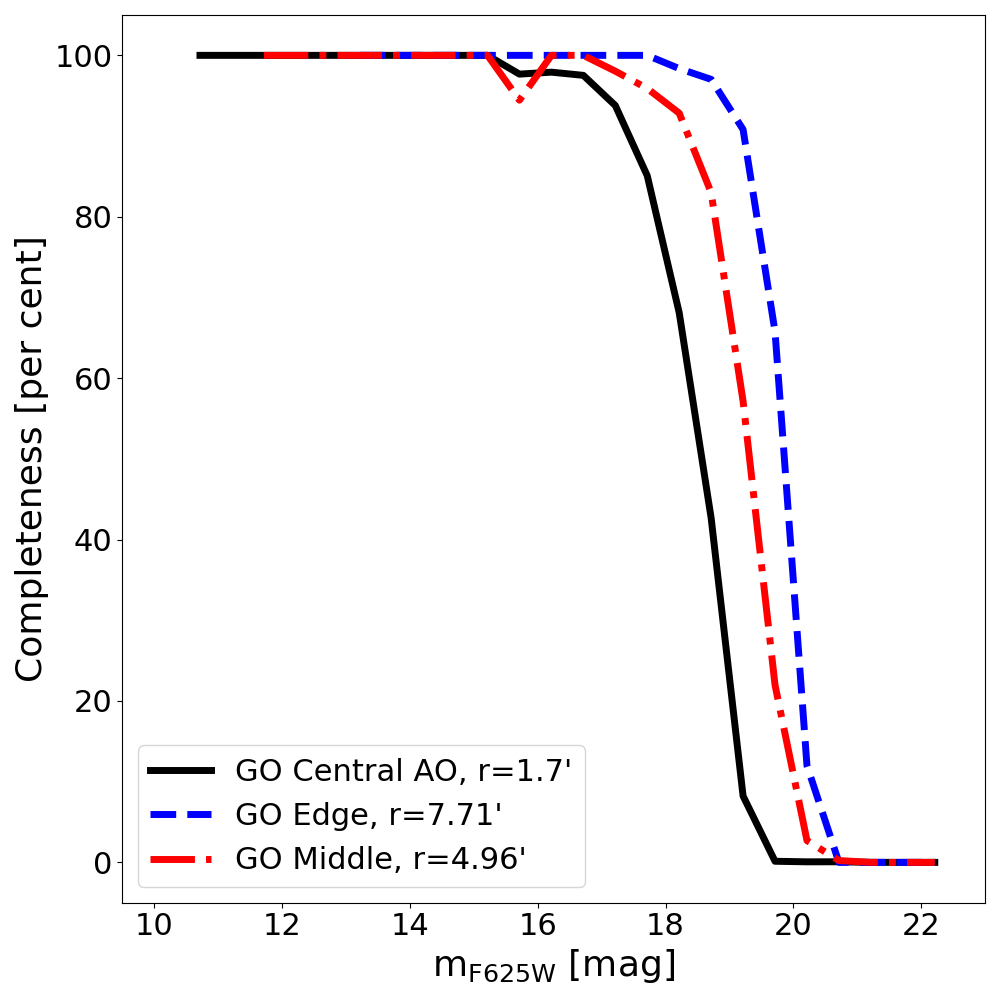}
\caption{\textbf{GO Completeness plots.} On the left is the overall GO completeness (the combined completeness for all 87 GO pointings) for different SNR cuts, from left to right the SNR is decreasing. The dashed line indicates 50 \% completeness and the dotted line is 18.86 mag in F625W. The right plot shows for an SNR $>$ 10 the individual completeness for 3 pointings at different radii from the center.\label{fig:comp_pointing}}
\end{figure*}

We investigate how complete our GO data are after the spectral extraction compared to the \textit{HST} catalog \citep{Anderson2010}.

For each MUSE pointing we mask the edges of the pointings by 5 arcsec, in order to not get edge effects, and we divide our sample in 0.5-mag-wide bins and define the completeness in each bin as the ratio between the number of stars in the \textit{HST} catalog and the number of stars for which we actually extract a spectrum from the MUSE data. We computed the completeness excluding spectra with an SNR lower than specific thresholds, i.e., 1, 5, 10, 15, 20, 25, and 30.

We show the overall completeness curves as a function of magnitude for different SNR cuts in the left panel of \autoref{fig:comp_pointing}. The completeness is almost 100 \% for the  brightest stars and falls off towards fainter stars. The left panel shows that the overall completeness is above 50 \% for stars brighter than $\sim$18.86 mag (F625W Filter) for an SNR threshold of 10, while for higher SNR this is shifted to lower magnitudes.  
The right panel shows that in the central regions, the completeness is lower for the fainter stars due to crowding.
\begin{figure}
    \plotone{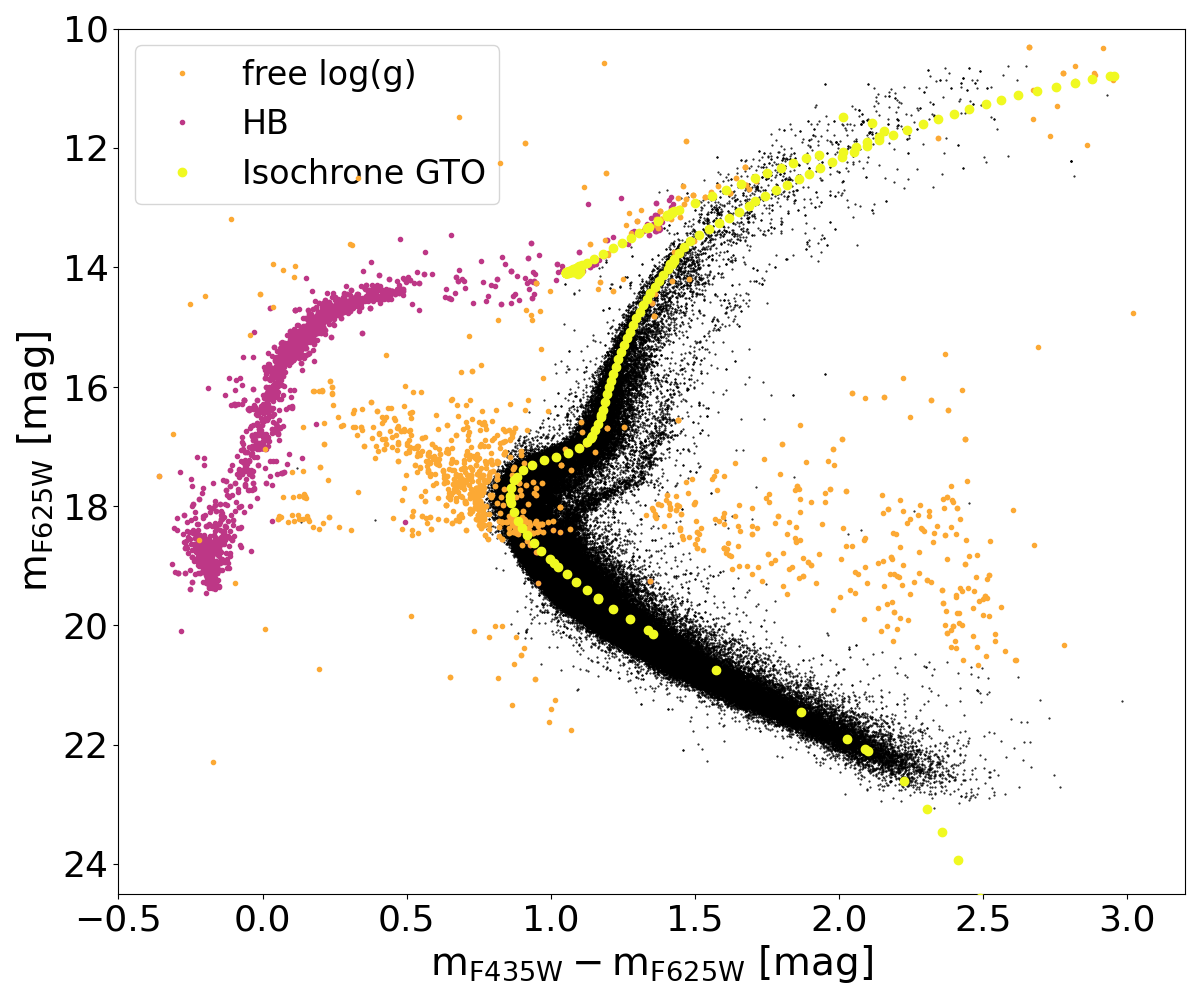}
    \caption{\textbf{CMD of GO stars.} CMD of the stars in the GO dataset. The black dots show the stars with fixed log(g) in our \textsc{spexxy} fit while the orange points have free log(g) and the purple are the HB stars which are disregarded in the current analysis. The yellow points indicate the position of the isochrone used to infer the log(g). \label{fig:CMD spexxy}}
\end{figure}

\begin{figure*}[h!]
\gridline{\fig{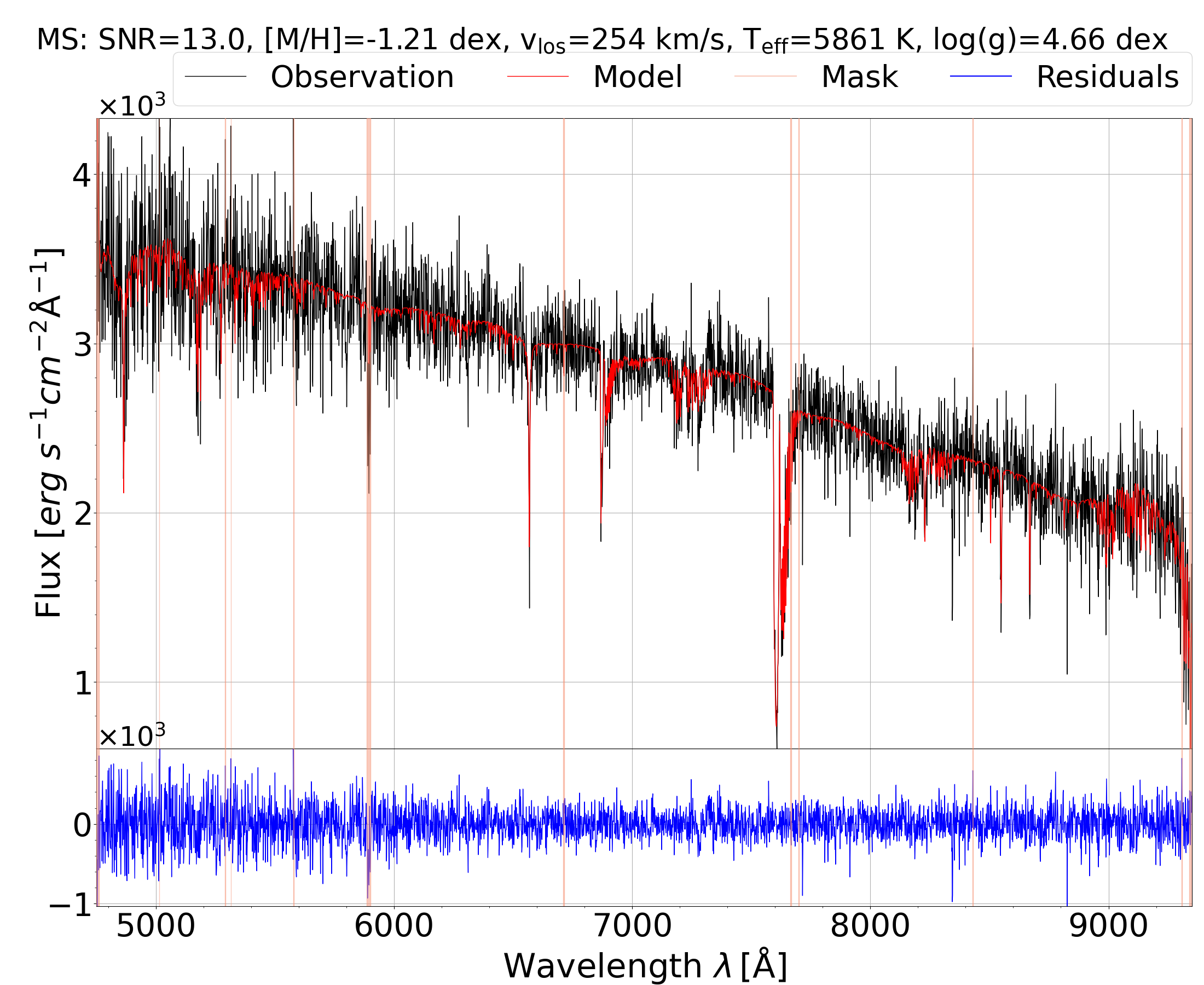}{0.44\textwidth}{(a) Main sequence star}
          \fig{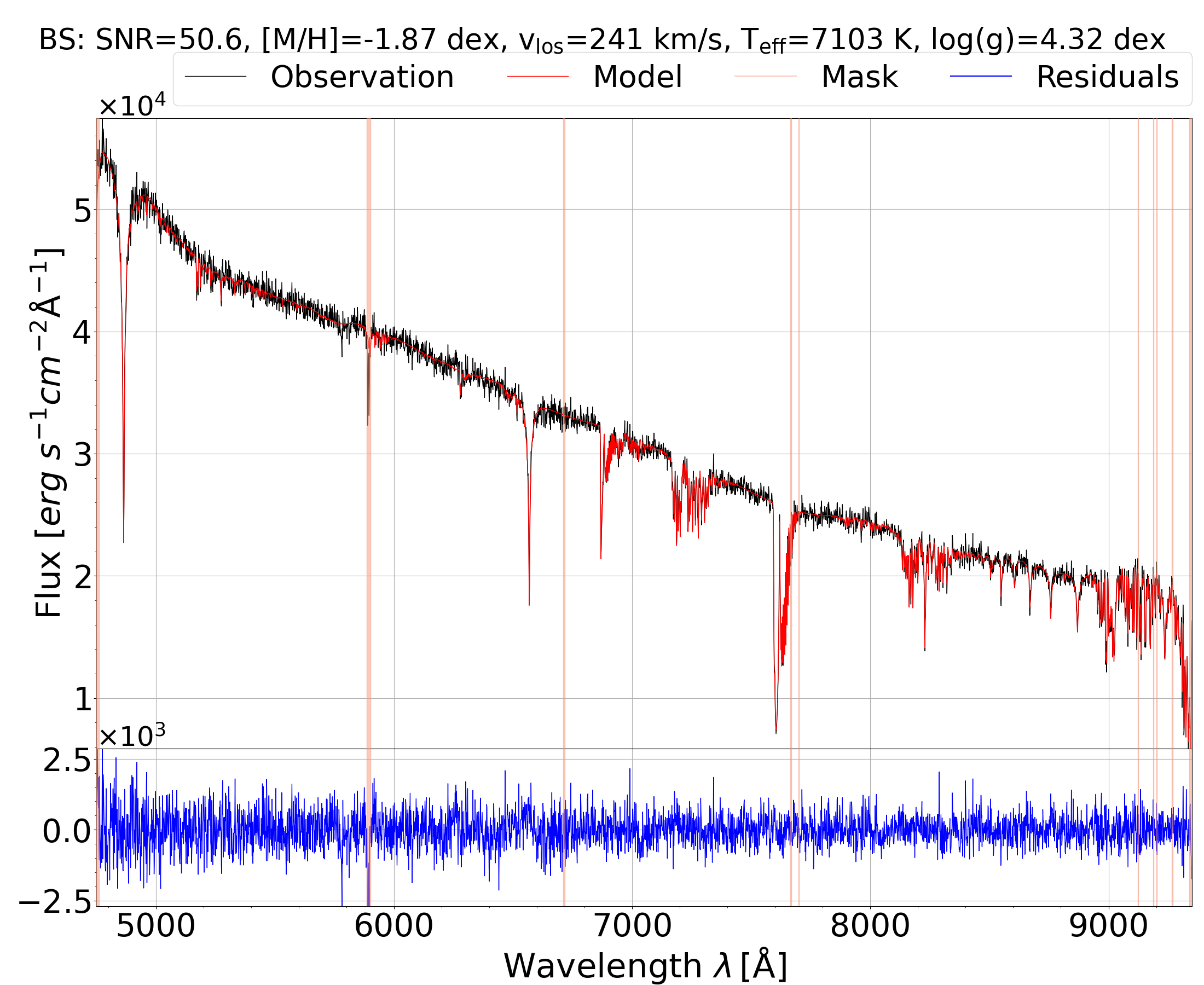}{0.44\textwidth}{(b) Blue straggler star}}
\gridline{\fig{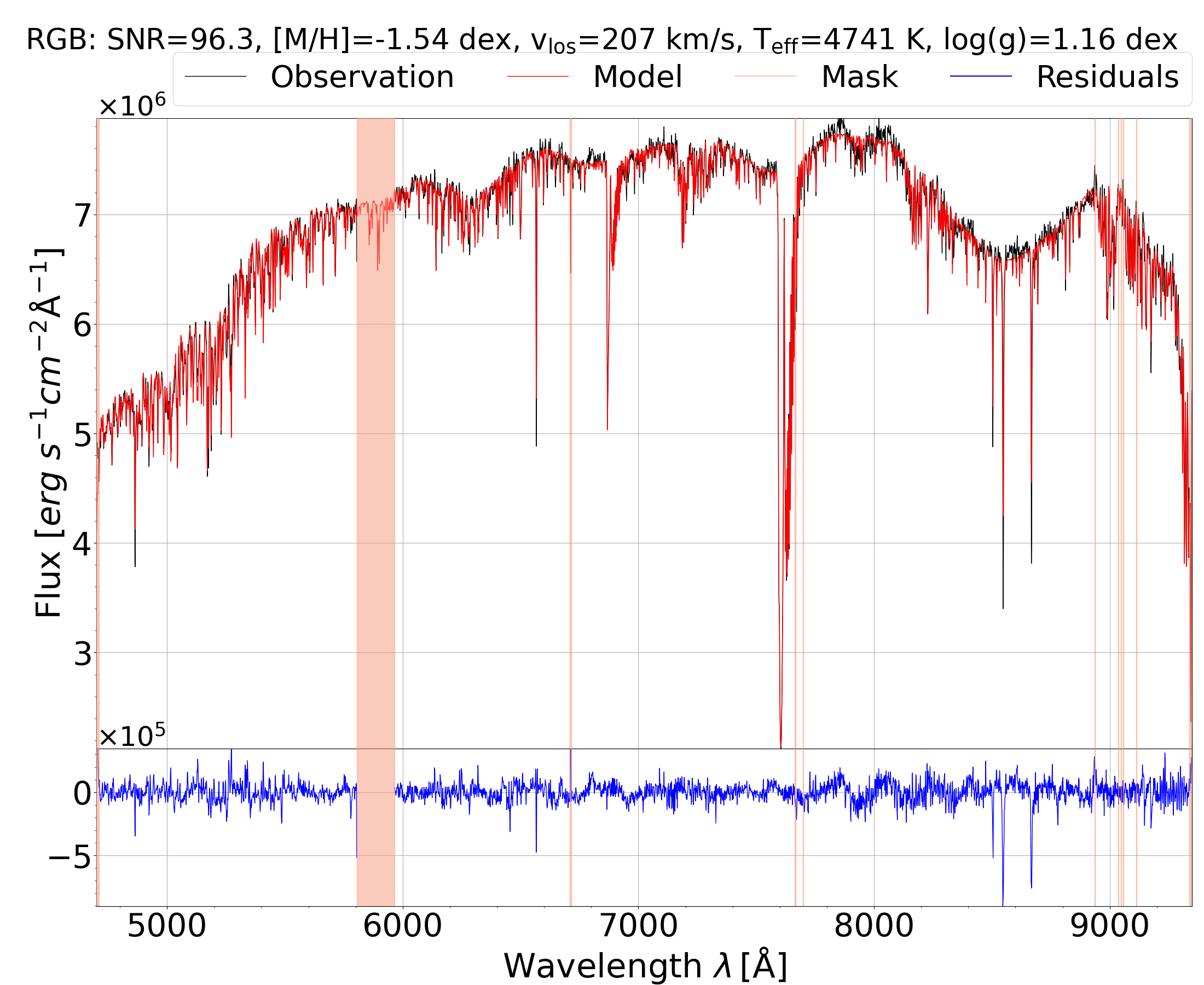}{0.44\textwidth}{(c) Red giant branch star}
          \fig{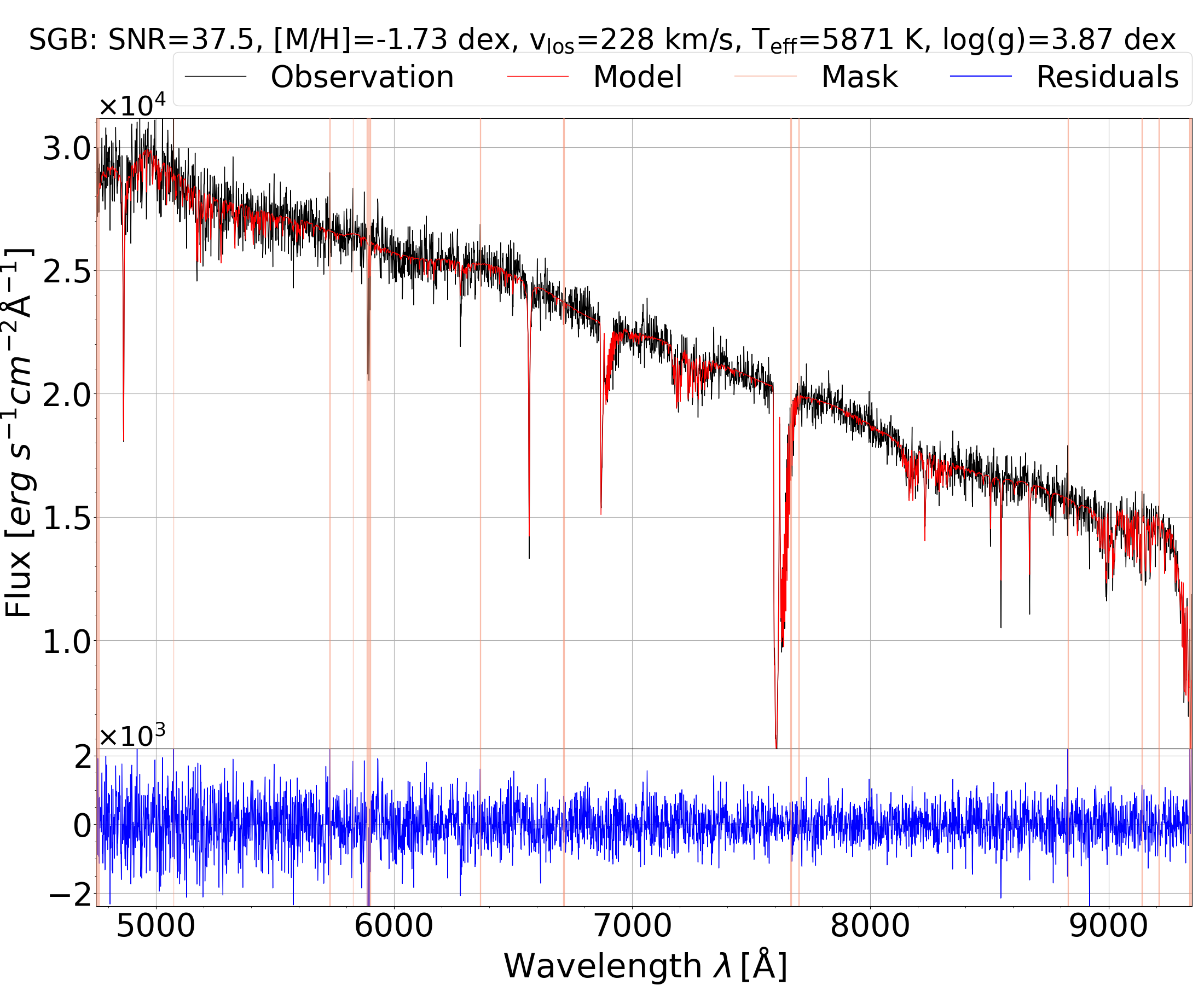}{0.44\textwidth}{(d) Subgiant branch star}}

\caption{\textbf{Example spectra.} (a)-(d); Spectra of stars at different stages of evolution plotted in black, with best-fitting \textsc{spexxy} models in red. The light red shaded areas are the excluded/masked regions during the fit; note that the RGB star (d) is observed with AO and the NaD line is blocked. The bottom panels show the data-model residuals in blue. We note that the deep broad absorption features at 6800 and 7600 \AA~are telluric features that are fit by \textsc{spexxy} along with the stellar properties. The best-fitting parameters for SNR, [M/H], $\rm v_{los}, T_{eff}$ and the log(g) value are given on top of each spectrum.} \label{fig:spectra}
\end{figure*}

\subsection{Spectral fitting}\label{sec:spexxy}

After extracting the spectra, we use \textsc{spexxy}\footnote{\url{https://github.com/thusser/spexxy}} version 2.5 to measure the physical parameters of the stars. The observed spectra are compared to synthetic spectra from the Phoenix library \citep{Husser_2013} to derive each star's effective temperature $\rm T_{eff}$, and metallicity [M/H] using a $\chi^2$ minimization. The range of the spectral parameters are: $-4$\,dex$\leq [M/H]\leq 1$\,dex, \SI{2300}{\kelvin}$\leq T\leq$ \SI{15000}{\kelvin} and $0\leq log(g)\leq 6$. We also include a \textsc{spexxy} fit to the telluric absorption lines (which were not corrected previously, see \autoref{sec:data red}) for $\rm H_{2}O$ and $\rm O_{2}$. We use all spectra we could extract in this analysis part regardless of their quality at this point.

The initial values for log(g) and $\rm T_{eff}$  were determined as outlined in \citet{Husser_2016, Kamann_2018}. In summary, the photometry from \citet{Anderson2010} was fitted with an isochrone from the PARSEC database \citep{Marigo_2017} and each observed star was assigned the stellar parameter from the nearest isochrone point, in the magnitude-color plane. For that, a two dimensional polynomial is fitted to the temperature and log(g) from the isochrone as a function of color and magnitude. This isochrone has an age of \SI{13.7}{\giga\yr}, a metallicity of -0.94\,dex, extinction ($A_{\rm v}$) of 0.23\,mag, a distance of \SI{5320}{\pc}, and is shown as yellow dots in \autoref{fig:CMD spexxy}. Since it is known that $\omega$~Cen has multiple populations, one isochrone is not ideal to describe all populations but here we only want to get initial values and a reliable log(g) for our spectral fit. A detailed population analysis will follow in upcoming work. The same is true for binary stars or multiple stellar systems, which are also offset in the color-magnitude diagram (CMD). However, the binary fraction of $\omega$~Cen is small, less than 5 \% \citep{Elson_1995, Mayor_1996} and in recent work even lower at $2.70\% \pm 0.08\%$ \citep{Bellini_2017} and because of multiple populations, the binary sequence is also intertwined with the main sequence track. Since log(g) does not strongly affect the [M/H] determination \citep[][see also \autoref{sec:logg free}]{Husser_2016, Kamann_2016} and we only require reliable log(g) parameters, one isochrone is sufficient even if the cluster is much more complex. We show the CMD in \autoref{fig:CMD spexxy}. During the fit with \textsc{spexxy}, log(g) is fixed to the value provided by the isochrone ranging from 0 to 6, the $\alpha$-enhancement is kept constant at [$\alpha$/Fe]=0.3\,dex \citep[similar to][]{Latour_2021}, while $\rm T_{eff}$ and [M/H] are determined. \textsc{spexxy} also measures the line-of-sight (LOS) velocity, $\rm v_{los}$, for the observed spectra. For that, we need initial values, which we get from a first cross-correlation with templates before starting the run with \textsc{spexxy}. If the result of the cross-correlation is not good enough, and we obtain  $r_{cc} \leq 4$ from the r-statistics \citep{Tonry1979}, we set the initial value to \SI{234}{\kilo\meter\per\second}, which is the mean LOS velocity of the whole cluster \citep{Baumgardt2019}.

For the spectral fit, we use the whole wavelength range of the MUSE spectra, excluding only the NaD line or AO region (\SI{578}{\nano\meter}~$-$~\SI{599}{\nano\meter}) and any other bad pixels identified. Four example spectra are shown in \autoref{fig:spectra} together with their best-fitting template spectrum in red. The light red shaded areas are the masked regions excluded from the fit. In addition, the SNR is also determined by \textsc{spexxy} from the fit residuals providing a good indicator of the quality of the data.

In summary, with \textsc{spexxy} we derive values for [M/H], $\rm v_{los}$, $\rm T_{eff}$, SNR, telluric parameters and their statistical errors. However, the results for stars lying further away from the isochrone (e.g. blue stragglers sequence (BSS) or asymptotic giant branch stars (AGB)) that is used to provide the log(g) values have to be considered biased. Hence, for stars not lying on the main isochrone, i.e. stars indicated with orange dots in \autoref{fig:CMD spexxy}, we rerun \textsc{spexxy} allowing the log(g) parameter to vary freely during the fit. These stars are either not following the main CMD track (this might be nonmember stars or have unreliable photometry measurements from \textit{HST}) or belong to the BSS or AGB. For stars that could be BSS/AGB or be part of the main CMD track, we keep both sets of values with free and fixed log(g). These are the orange dots that lie on top of black dots in \autoref{fig:CMD spexxy}. This increases the number to 356,065 measurements due to the duplicates added from the overlap regions. 

For the further analysis, we do not include the horizontal branch (HB) stars, indicated as  purple dots in \autoref{fig:CMD spexxy}. Stars hotter than $\sim$11,000~K have chemically peculiar atmospheres due to diffusion and require dedicated model atmospheres for proper analysis \citep[see][]{Latour_2023}. The analysis of the HB stars included in our GO data will be presented elsewhere.

\section{Analysis}\label{sec:results}

In \autoref{sec:methods} we explained how we get a spectroscopic catalog from our MUSE data. This catalog needs to be further analyzed to remove contamination from foreground/background stars and to obtain robust measurements. Our overall goal is to keep as many stars as possible in the catalog and provide all parameters needed to perform the relevant quality cuts depending on the science case. 

For our final catalog, we combine both the GO and GTO datasets. The GTO catalog contains 795,944 measurements belonging to 75,416 unique stars (see \autoref{sec:multi}). As described in the previous section, the GO analysis followed that of the GTO data with three main differences: (1) the GTO data have multiple epochs of observations as opposed to a single epoch for the GO data, (2) the GTO use two \textit{HST} catalogs for source positions, the catalog created by \citet{Anderson2008} for the ACS survey of Galactic GCs \citep{Sarajedini2007} and the \citet{Anderson2010} catalog, and (3) during \textsc{spexxy} fits the log(g) for the BSS in the GTO data set was kept fixed to 4.2 dex, while we fit for log(g) for these stars.

\subsection{Reliability parameter}\label{sec:reliability}

To have one overall parameter evaluating the quality of the derived LOS velocities, we calculate a reliability parameter $R$, as explained in Sect. 3.2 of \citet{Giesers2019}:
\begin{eqnarray}
    R = (2R_{SNR} + 10R_{cc} + R_{\epsilon_{vcc}} + 3R_{v} + 2R_{\epsilon_{v}}  \nonumber\\ + 5R_{v=vcc} )/23.
\end{eqnarray}
The different components can be either false (0) or true (1), allowing the range of $R$ to be from 0 to 1 with higher values being more reliable than lower ones. The first component, $R_{SNR}$, is a SNR cut, while the other parameters are different checks of the credibility of the LOS velocity calculations. In detail, they check the quality of the cross-correlation ($R_{cc}$), if the cross-correlation and the full-spectrum fit uncertainty are plausible ($R_{\epsilon_{vcc}}$ \& $R_{\epsilon_{v}}$), if the full-spectrum fit velocity is agreeing with the expected cluster or foreground velocity and with the cross-correlation ($R_{v}$ \& $R_{v=vcc}$).

First, we consider a minimum SNR of 5 to be reliable:
\begin{equation}
R_\mathrm{SNR}=\mathrm{SNR}\geq 5.
\end{equation}
That is however not our suggested SNR cut for the whole catalog, which should be higher (\autoref{sec:catalog}), but only for the velocity calculation which does not need that high SNR. In addition, we use the FWHM result from our cross-correlation and the r-statistics \citep{Tonry1979} to check the quality of the cross-correlation which gave the initial LOS velocities for the fit with \textsc{spexxy}:
\begin{equation}
R_{cc} = r_{cc}\geq 4 ~ \& ~FWHM > 10 \AA.
\end{equation}
Further, a plausible velocity uncertainty of \SI{0.1}{\kilo\meter\per\second} for the velocity from \textsc{spexxy} and from the cross correlation is required:
\begin{eqnarray}
R_{\epsilon_{v}}=\epsilon_{v}> \SI{0.1}{\kilo\meter\per\second} \\
R_{\epsilon_{vcc}}= \epsilon_{vcc}> \SI{0.1}{\kilo\meter\per\second}.
\end{eqnarray}

Additionally, the velocity output from \textsc{spexxy} should be plausible. This means that it should match the velocity of either the cluster or the Galactic field stars. Hence, the values are checked to be within 3$\sigma$ of the cluster velocity and the cluster dispersion, $v_{cl}$ = (232.7 $\pm $0.2)\si{\kilo\meter\per\second} and $\sigma_{cl}$ = \SI{17.6}{\kilo\meter\per\second} \citep{Baumgardt2018}, or that they agree with the foreground stars. For the cluster membership, we calculate:
\begin{equation}
    R_{v_{1}}=\frac{|v-v_{cl}|}{\sqrt{\epsilon_{v}^2 + (\sigma_{cl} + \epsilon_{v_{cl}})^2 + a^2}}\leq 3.
\end{equation}
With $a$ being \SI{30}{\kilo\meter\per\second} which is added to the broadening in order to account for the orbital motions of binary stars, as their measured velocities
would not necessarily be covered by the cluster distribution. For the foreground stars:
\begin{equation}
    R_{v_2}=\frac{|v-0km~s^{-1}|}{\sqrt{\epsilon_{v}^2 + (50 km~s^{-1})^2}}\leq 3,
\end{equation}
with \SI{50}{\kilo\meter\per\second} being the estimate for the foreground stars from the Besan\c{c}on model \citep{Robin_2003} used in \autoref{sec:members}.
The total $R_{v}$ is true if at least one of the $R_{v_{1,2}}$ is true. Similarly, a maximum 3$\sigma$ difference is allowed between the cross-correlation results and the final \textsc{spexxy} results, while allowing a maximal velocity error of \SI{40}{\kilo\meter\per\second} for the cross-correlation:
\begin{equation}
R_{v=vcc}=\frac{|v-v_{cc}|}{\sqrt{\epsilon_{v}^2 + \epsilon_{cc}^2}}\leq 3.
\end{equation}

We include the total $R$ parameter for all stars in the catalog.

\subsection{Error Analysis}\label{sec:errors}

\begin{deluxetable*}{lcccccc}
\tablewidth{2.0\columnwidth}
\tablenum{1}
\tablecaption{Error Analysis \label{tab:error analysis}}
\tablehead{
\colhead{dataset} & \colhead{a} & \colhead{b}& \colhead{Maximum scaling} & \colhead{Minimum scaling} & \colhead{Maximum} & \colhead{Minimal}\\
\colhead{Parameter} & \colhead{} & \colhead{}& \colhead{Model/Data} & \colhead{Model/Data} & \colhead{fractional residual} & \colhead{fractional residual}}
\startdata 
GO [M/H] & 0.887 & 0.007 & 1.644/1.938 &1.401/1.200 & 0.179 & 0.226\\
GO $\rm v_{los}$ & -0.143 & 0.062 &  2.819/2.943 & 1.229/1.351 & 0.243 & 0.186\\
GO $\rm T_{eff}$ & 1.791 & -0.003 & 1.639/1.701 & 1.292/1.213 & 0.279 & 0.169\\
GTO [M/H] & 0.866 & -0.002 & 1.359/1.402 & 1.293/1.246 & 0.071 & 0.051\\
GTO $\rm v_{los}$ & -0.137 & 0.009 & 1.354/1.346 & 0.979/0.982 & 0.022 &  0.021\\
GTO $\rm T_{eff}$ & 1.463 & -0.014 & 1.521/1.596 & 0.976/1.137 & 0.209 &  0.125\\
\enddata
\tablecomments{Column 1: The dataset and parameter used for the error analysis; column 2, and column 3, are the best fitting parameters for the first order polynomial, g(x) = a + b$\cdot$ x, where x the SNR and y the $\sigma^{2}(\Delta y$). Columns 4 and 5 are the maximum/minimum scaling we get from our best fit and what the actual data have; columns 6 and 7 are the Maximum and Minimum fractional residuals, which are always under 30 \%.}
\end{deluxetable*}

\begin{figure*}[t]
\plottwo{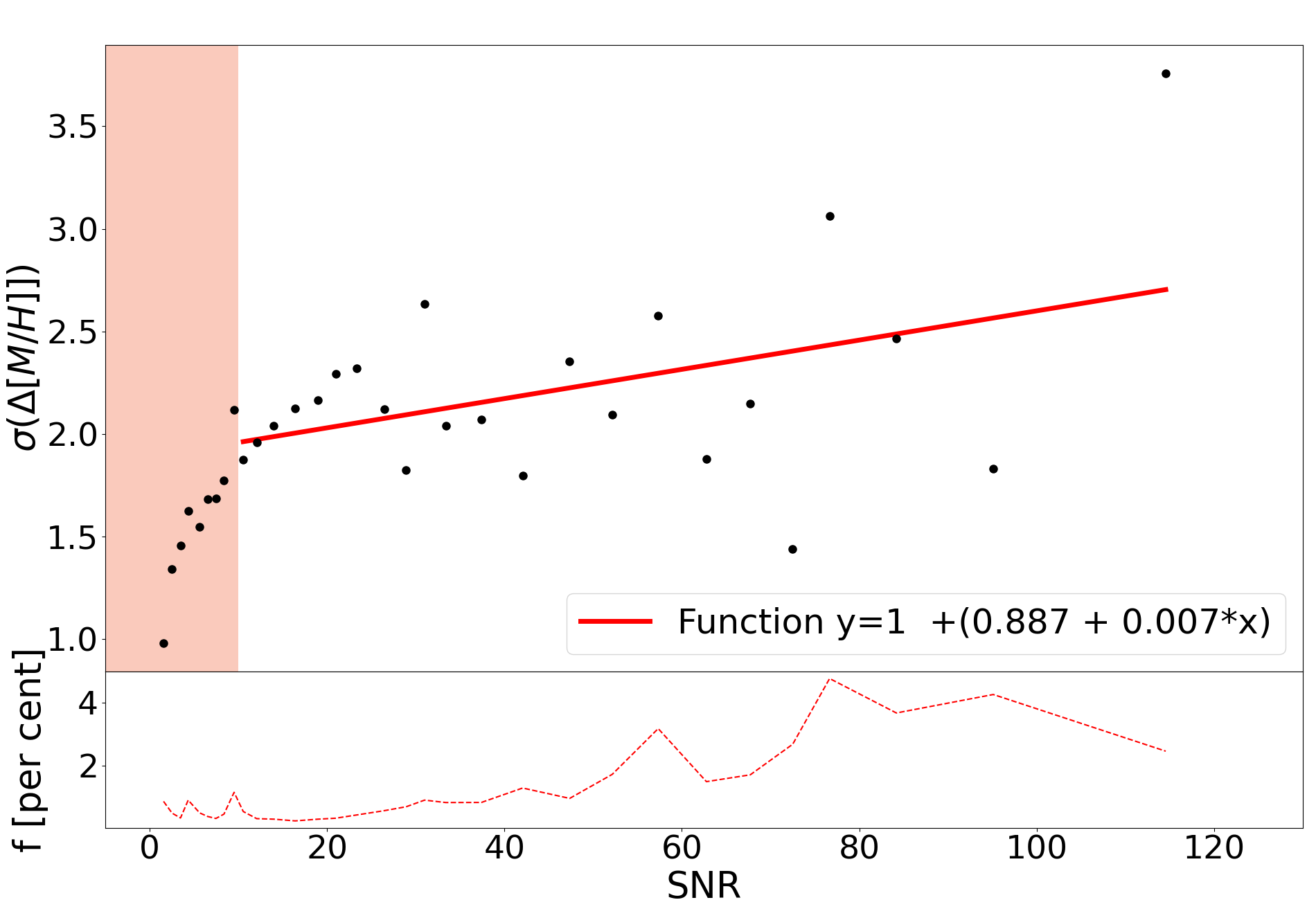}{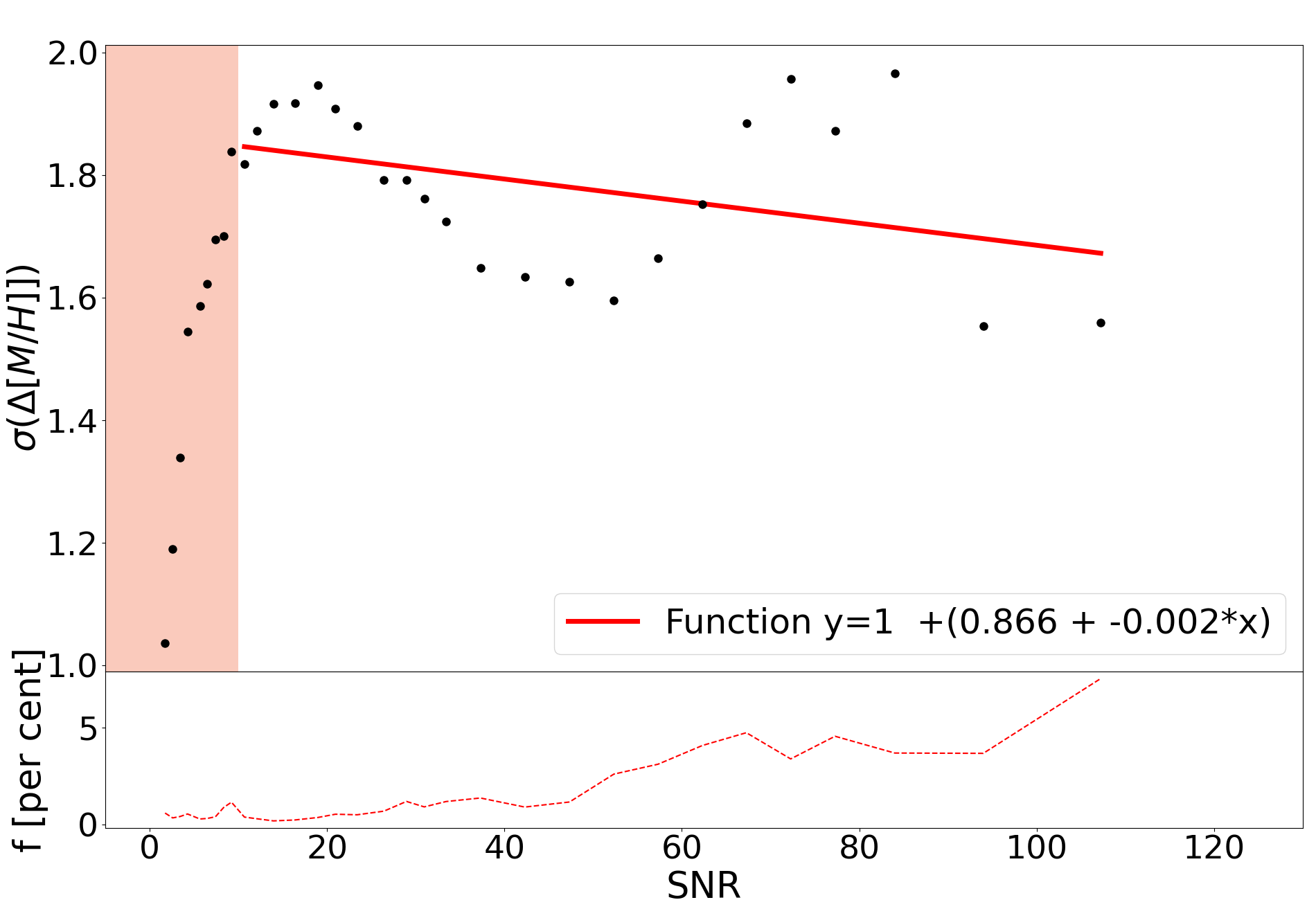}
\caption{\textbf{Error analysis correction.} Top panels: The x-axis is the SNR and the y-axis is the $\sigma^2$ of the $\Delta [M/H]$ for the GO data on the left and the GTO data on the right. The red line is the best-fitting first-order polynomial, the black dots are our data values, and the light red shaded area is the region below an SNR of 10, which we exclude from the analysis. Bottom panels: The x-axis is the SNR and the maximum percentage of counts one star has in each bin. The red dashed line shows the maximum number of measurements each star can have, which is clearly always below 10~\% of the total number of measurements in each bin, making sure the statistics in each bin are not the result of only one star.\label{fig:error SN}}
\end{figure*}
\begin{figure}[]
\plotone{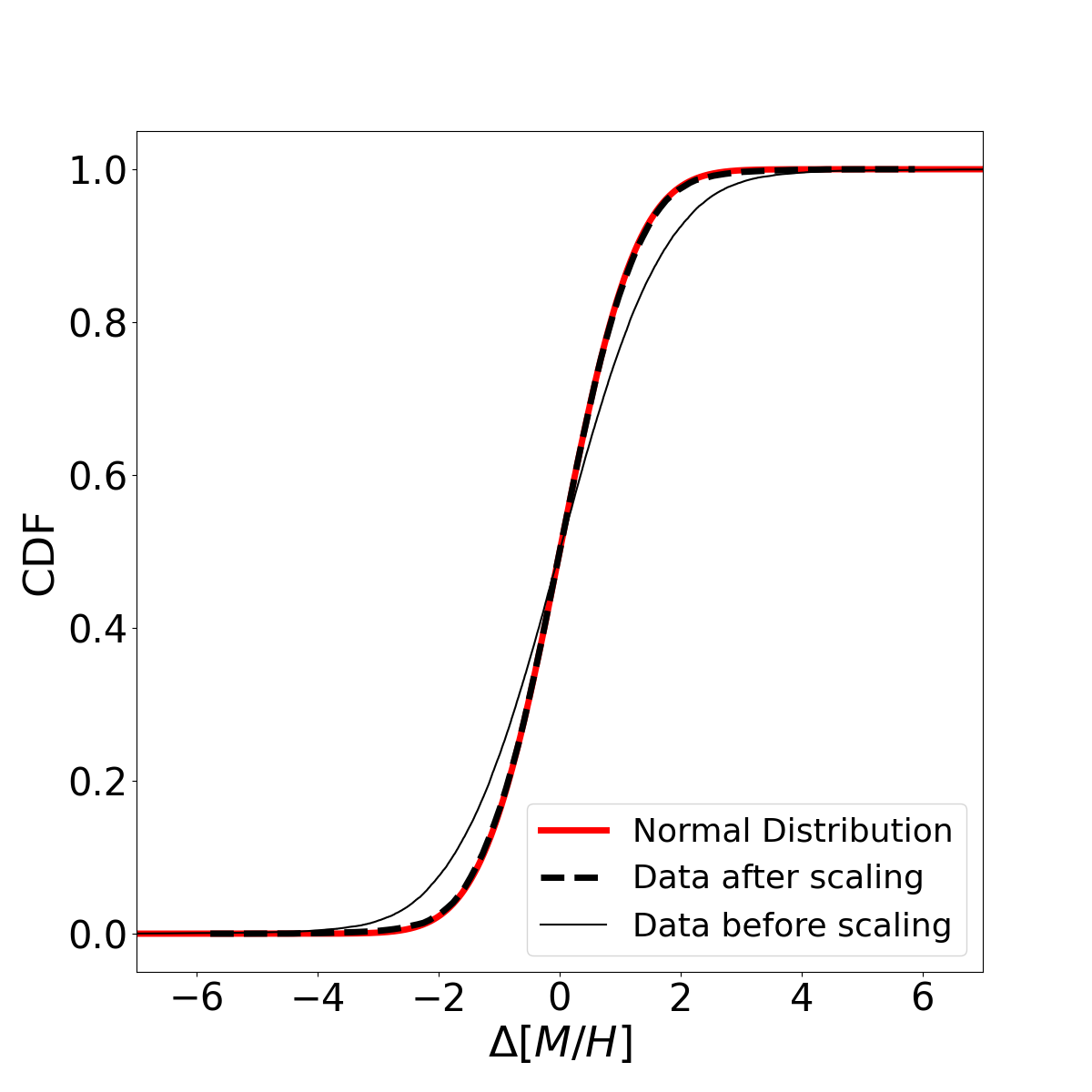}
\caption{\textbf{Error analysis CDF.} The cumulative distribution of the $\Delta [M/H]$ in black for SNR $>10$ for the GO data and in red the distribution if it was a normal distribution. The dashed black line is the distribution after the error correction.\label{fig:error cdf feh}}
\end{figure}

To get a better understanding and estimate of the errors on our fitted parameters, we use stars with multiple observations (see \autoref{sec:multi}), allowing us to compare the different results we get for the same star, similar to \citet{Husser_2016, Kamann_2018}. From that, we can compare the statistical uncertainty we get from the spectral fitting routine with the scatter from the repeat measurements. This allows us to estimate how accurate the errors from the spectral fitting routine are, and any scaling factor needed to correct them. We do this analysis for our main results from \textsc{spexxy}: the metallicity, line-of-sight velocity, and effective temperature, using only the individual measurements used to compute the mean parameters, as described in the following section (\autoref{sec:multi}), for the error analysis. 

We calculate:
\begin{equation}
    \Delta y = \frac{y_{i} - y_{j}}{\sqrt{\epsilon_{i}^{2} + \epsilon_{j}^{2}}}, i \neq j 
\end{equation}
where $y_{i}, y_{j}$ are the different results we get from \textsc{spexxy} (where $y$ = [M/H], $\rm v_{los}$ or $\rm T_{eff}$) for the same star in two different observations, and $\epsilon_{i}, \epsilon_{j}$ are their \textsc{spexxy} errors. Some stars are in more than two observations (maximum in 4 for the GO data, 66 for the GTO data) so we calculate $\Delta y$ for each combination of different results, giving us 24,216 values for the GO dataset and 2,016,633 for the GTO.

Afterwards, we bin all $\Delta y$ values in SNR bins depending on the mean SNR of the two measurements, $SN_{i}$ and $SN_{j}$. We defined the SNR bins so that for the GO and GTO data, the statistics of no SNR bin can be dominated by just one star with multiple measurements if placed in the same bin. In the bottom panel of \autoref{fig:error SN} we show the maximum percentage of counts (f = max[counts one star]/total number counts, per bin) one star has in each bin, whereas for the GO data it is always below 5~\% and for the GTO below 10~\%. The distributions of $\Delta y$ in each of the SNR bins, if the \textsc{spexxy} errors are accurately estimated, should be a Gaussian with $\sigma=1$ and the mean at 0.

In \autoref{fig:error cdf feh} we show an example of the cumulative distribution for the metallicity errors before any correction (black solid line) in comparison to a normal distribution (red line). We find that, as in this example for SNR $>$ 10, the \textsc{spexxy} errors typically are underestimated relative to the repeat measurements, which are a better estimate of the true errors. We, therefore, multiply the \textsc{spexxy} errors by a scaling factor, $\rm \sqrt{1 +g(SNR)}$, to get more accurate errors for each star. After applying the scaling of the errors (see next paragraph), the cumulative distribution of the repeat measurement $\Delta$[M/H] (black dashed line in \autoref{fig:error cdf feh}) follows the normal distribution.

If we plot the variance $\sigma^2$ ($\sigma$ calculated as half of the difference between the 15.86 percentile and 84.14 percentile) of this distribution as a function of SNR, we can see that $\sigma^2$ is higher than 1 towards larger SNR, see \autoref{fig:error SN}. We can describe this trend using a simple function, y = 1 + g(x), with g(x) being a linear function, in order to get an estimate of the error correction needed as a function of SNR. We exclude the SNR bins below 10 for the fitting of the line. In \autoref{tab:error analysis} we list the best fitting parameters for the scaling relation and the values for the maximum and minimum scaling for SNR $>$ 10. To quantify how accurately the linear function captures the repeat measurement data, we also include the maximum/minimum fractional residual to this fit (i.e.~the fractional error on the point that lies farthest from the line in \autoref{fig:error SN}). 

The derived scaling functions are then applied to all data according to their SNR. The mean scaled errors for 40 $<$ SNR $<$60 are $\epsilon_{v_{los}}$ = \SI{3.11}{\kilo\meter\per\second}, $\epsilon_{T_{eff}}$ = \SI{36}{\kelvin} and $\epsilon_{M/H}$ = 0.057~dex. These new, scaled errors are more accurate estimates of the true errors and hence we will always use them in the following analysis.

\subsection{Combining multiple measurements}\label{sec:multi}

The next step is to combine measurements for each observed star. Due to the multiple epochs of observations in the GTO data, each star has been observed multiple times, and the spectra were analyzed independently, such that the 795,944 measurements belong to 75,416 unique stars. For the GO data, only stars in the overlap regions of the different pointings (green and yellow regions in \autoref{fig:data}) have been observed more than once, with 303,822 unique stars out of the 356,065 measurements. In addition, there are also overlapping regions in the GTO and GO data where stars are in both datasets.

To combine multiple measurements of stars, we do a mean calculation for both datasets together after scaling the errors. First, we require the measurements to have a successful \textsc{spexxy} fit, meaning a fit to the templates could be performed without problems and the final parameters are not close to their limits set for the fit. These cuts leave us with 744,344 GTO measurements and 335,834 GO measurements. 

For stars with multiple measurements, we then calculate the mean values using a subset of measurements selected in a procedure based on their reliability value (\autoref{sec:reliability}), magnitude accuracy from the spectral extraction, the distance to the edge of the detector, and an SNR cut. We use a threshold reliability value of 0.5, and magnitude accuracy (the relative accuracy of the magnitude from the extracted spectrum vs. the \textit{HST} catalog) $<$0.6. We continue with measurements that meet these thresholds, or if there are none for a single star, we keep all measurements for the next steps in the combination. 

Next, to use comparable measurements in the mean calculation, we first find the highest SNR spectrum and check if it is at least \SI{5}{\pixel} (1\arcsec) away from the edge of the pointing. If it is, we consider it the best measurement for the given star. If it is not, we check the second-highest SNR spectrum and set that as the best measurement or, if it is also too close to the edge, we return to the highest SNR spectrum as the best measurement. Having found the best measurement, we require all remaining measurements to have at least half the SNR of the best measurement or in general higher than 20, at least as good of an extraction quality flag as the best, and if the best measurement was more than \SI{5}{\pixel} away from the edge, the other measurements have to be, too. The measurements that pass these tests are in general comparable to the best measurement and are used to calculate the mean measurements for a given star. If there is only one measurement, then that is used as a single entry for that star. 

For the mean calculation of [M/H], $\rm v_{los}$ and $\rm T_{eff}$ we calculate the inverse-variance weighted mean, using the scaled errors (see \autoref{sec:errors}):
\begin{eqnarray}
\hat{y} = \frac{\Sigma_i y_i/\epsilon_{si}^2}{\Sigma_i 1/\epsilon_{si}^2} \\
\epsilon_{s\hat{y}} = \sqrt{\frac{1}{\Sigma_{i} 1/\epsilon_{si}^2}}.
\end{eqnarray}
$\epsilon_{s\hat{y}}$ is the error to the inverse-variance weighted mean, similarly we calculate the combined not scaled \textsc{spexxy} errors. The scaling factor and SNR are normal unweighted mean values, while for the quality parameters: distance to the edge, magnitude accuracy, extraction quality flag, and reliability parameter, we keep the minimum value (maximum for the extraction quality flag) included in the combined value. For the log(g) parameter we give the unweighted mean value, indicating if it was kept fixed or free during the fit, if we used both free/fixed we give both means.

This leaves us with a combined catalog of 342,796 unique individual stars with physical parameters. We describe recommended quality cuts for using this catalog further in \autoref{sec:catalog}. We match this catalog with the \textit{HST} catalog by \citet{Anderson2010}, but note that 1,432 stars in the GTO data which were extracted using the \citet{Anderson2008} catalog could not be matched; these stars are included in the catalog without \citet{Anderson2010} magnitudes.

\subsection{Membership}\label{sec:members}

\begin{figure}[t]
\plotone{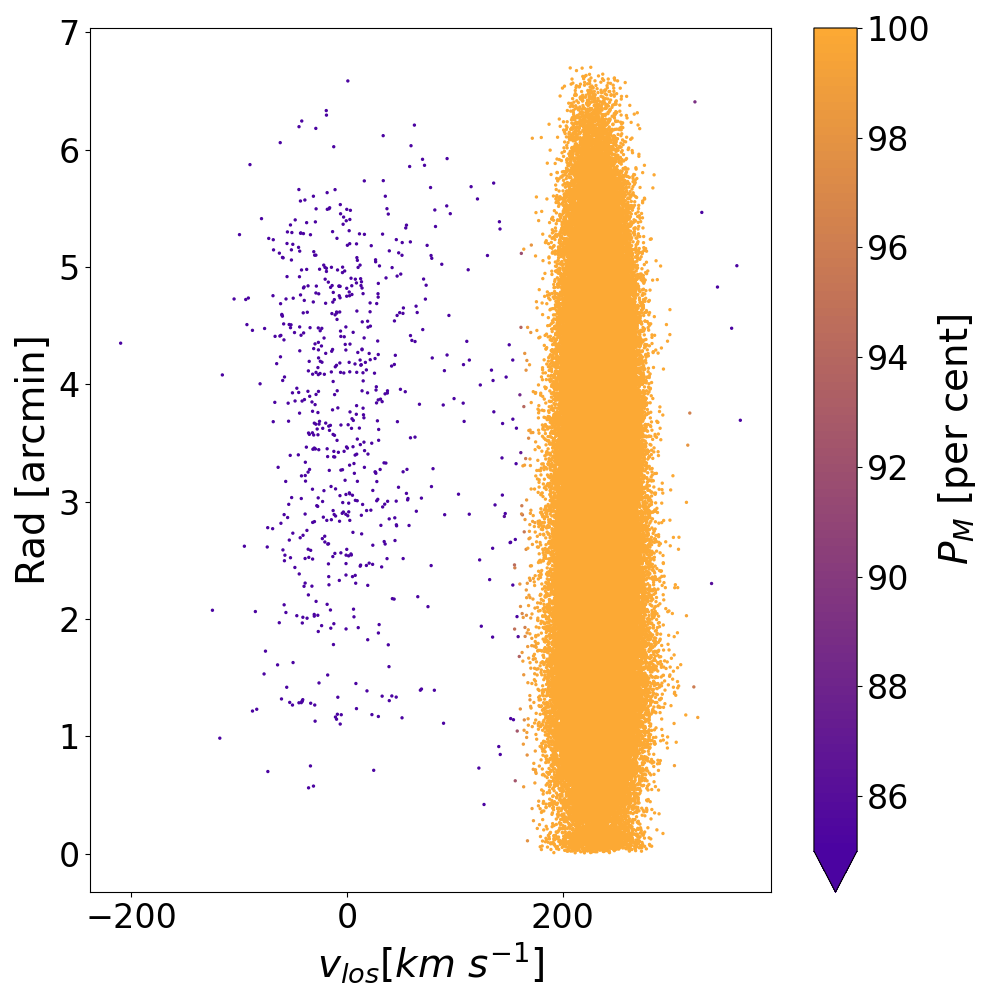}
\caption{\textbf{Membership probabilities using radius and velocity for a SNR$>$10.} The stars' radial distances to the cluster center in arcminutes are plotted against their line-of-sight velocities. The points are color-coded according to their membership probability, yellow being 1.
\label{fig:membe1}}
\end{figure}

\begin{figure*}[t]
\plotone{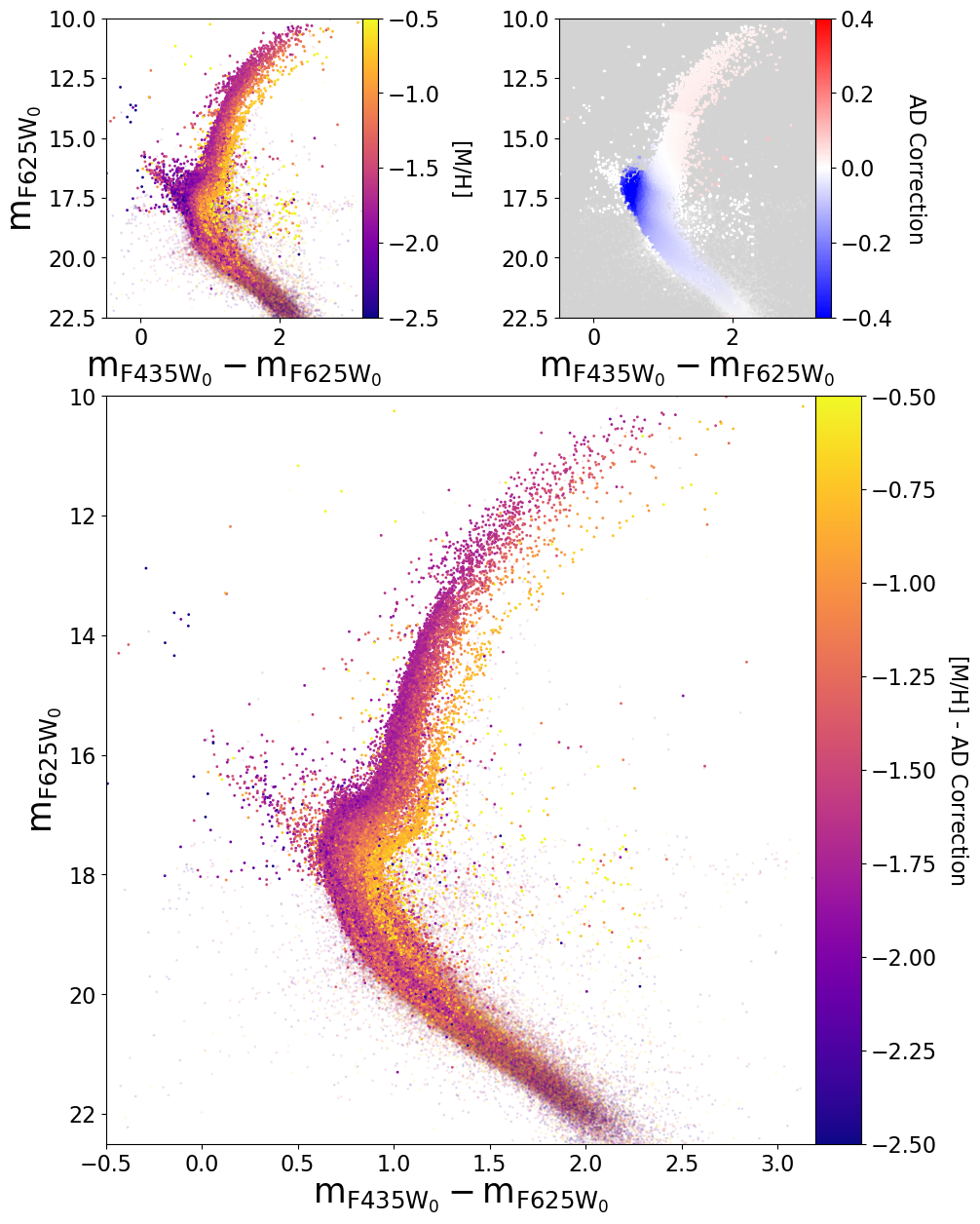}
\caption{\textbf{Atomic diffusion correction.} Top left: CMD of all stars with \textit{HST} photometry colored by their metallicity. Top right: A grid of the median metallicity correction when considering isochrones that model atomic diffusion. Bottom: CMD of all stars from the top left panel, now with the relevant atomic diffusion correction applied. Stars with $\rm SNR < 10$ and less reliable photometry are plotted with transparency in each panel. \label{fig:ad_corr}}
\end{figure*}

To decide which stars likely belong to $\omega$~Cen, we assign the stars a membership probability $P_{M}$ that we can use to exclude foreground or background stars belonging to the Milky Way. We determine this probability using the python package \textsc{clumpy}\footnote{\url{https://github.com/bkimmig/clumpy}} \citep{Kimmig_2015} which is based upon \citet{Walker_2009}.
This code can use either radius and velocity only for the estimate or can include another parameter, like the metallicity, to get the membership probability. As a foreground velocity distribution, we use the Besan\c{c}on model \citep{Robin_2003}\footnote{\url{https://model.obs-besancon.fr/ws/}} with 1 degree on each side centered on $\omega$~Cen and the model maximum absolute velocities are reaching up to \SI{500}{\kilo\meter\per\second}.

The package uses an iterative maximization technique to get the probability and the systemic velocity and dispersion. Three separate memberships are going into the total membership probability including information about radius, velocity, and the foreground model: i) the probability of being a member which assumes that the stars follow a Gaussian velocity distribution centered on the mean velocity of the cluster, ii) the non-membership probability including the Besan\c{c}on model information assuming a velocity dispersion of \SI{20}{\kilo\meter\per\second}, and iii) the probability depending on the radius assuming a radial decrease in membership. After every iteration, the mean velocity, velocity dispersion, and probability are updated until they converge. We perform 50 iterations and the initial values are 0.5 for the probabilities and \SI{4}{\kilo\meter\per\second} for the central dispersion of the cluster.

The membership output for SNR $>$ 10 is shown in \autoref{fig:membe1} when using only velocity and radius from the center \citep[$\mathrm RA_c$ = 13:26:47.24 and $\mathrm Dec_c$ =-47:28:46.45,][]{Anderson2010} as input. We clearly can see the separation between the foreground stars and the members (over 95 \% probability) in the velocity radius plane, with 98.8\% (338,531 stars) of stars being members and only 1.2\% (4,266 stars) not. We choose to use the membership that does not include metallicity information for the following analysis since the [M/H] distribution of the cluster is too broad and hence not that easily distinguishable from the foreground. We note that 117 stars have no membership probability since their velocity exceeds the \SI{500}{\kilo\meter\per\second} from the foreground model, however, these stars have low SNR ($\leq 5$).

\subsection{Atomic Diffusion Correction}\label{sec:atomic_diffusion}

In the left panel of \autoref{fig:ad_corr} we show the CMD of our full MUSE catalog with each point colored by its metallicity [M/H]. The stars near the main sequence turn-off (MSTO) seem to be more metal-poor than stars falling directly above and below this region. A similar offset in metallicity was seen by \cite{Husser_2016} in NGC~6397, who find a $\sim$0.3 dex variation in the mean metallicity along the MS, with the lowest metallicity at the turn-off. 
Other analyses have also found a decrease in the observed metallicity \citep[up to 0.25 dex,][]{King1998} (in M92/NGC 6341) compared to stars of the same population along the MS and subgiant branch (SGB). 

This shift can be explained by several internal transport processes including atomic diffusion (gravitational settling of heavier elements), thermal diffusion, radiative acceleration, and turbulent mixing \citep{VandenBerg2002}. For stars evolving through the MSTO, heavier elements begin to sink below the outer layers of the star causing them to be undetectable via spectroscopy. The consequence is that the observed metallicity of these stars is lower than the abundance they were originally born with. Once a star enters the SGB, convection in the outer layers begins and heavier elements are then dredged up to the surface where they are once again observable. Therefore, these stars no longer exhibit the offset in metallicity attributed to atomic diffusion. It is also noted in \cite{Korn2007} and \cite{Nordlander2012} that this phenomenon has a larger effect on extremely metal-poor stars, which is where we also see the greatest offset.  We note that hereafter we refer to our correction for these effects as an "atomic diffusion correction."
\begin{figure}[b]
\plotone{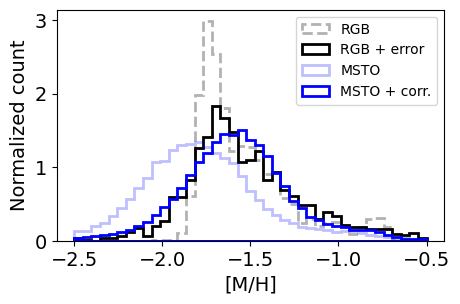}
\caption{\textbf{[M/H] of RGB vs. MSTO post AD correction.} The figure shows how the atomic diffusion correction causes the [M/H] distribution to be more consistent between the MSTO and the RGB (when considering both to have MSTO-like errors). \label{fig:adc_hist}}
\end{figure} 
To correct for this effect on our metallicity measurements, we use MIST isochrones (\citealp{MISTDotter, MISTChoi}), which take into account atomic diffusion. We use the distance modulus for the isochrones and apply extinction and reddening corrections to our photometry. Specifically, we assume  $ \rm A_v$ = 0.372 mag \citep{Harris1996, Bellini_2017a} and use the  values for $ \rm A_{\lambda} / A_{V}$ of  1.34148 and 0.85528 for F435W and F625W respectively \footnote{Obtained from \url{http://stev.oapd.inaf.it/cgi-bin/cmd},  \cite{Girardi_2008}.}, giving us $ \rm A_{\lambda, 435} = 0.499$ mag and $ \rm A_{\lambda,625} = 0.318$ mag. We then subtract $ \rm A_{\lambda}$ from the relevant magnitudes to obtain our corrected photometry. 

We use a series of isochrones with $\rm Age = 10 \ Gyr$ and $\rm -4.0 < \langle [Fe/H] \rangle < 0.5$. We increase the CMD coverage of the MIST isochrones (which are available in steps of 0.5 to 0.25 dex in metallicity) by interpolating between equivalent evolutionary points to generate isochrones between the given models. We then iterate this process until we obtain good coverage across the CMD and a metallicity precision of 0.03 to 0.06 dex. For each interpolated isochrone, we calculate the difference between its overall metallicity and the surface metallicity at each point along it. We grid these differences in the CMD using color bins of 0.05 mags and F625W bins of 0.12 mags, finding the median correction in each bin. 
To derive the atomic diffusion (AD) correction for individual stars, we interpolate within this 2D grid of corrections. The bottom panel of \autoref{fig:ad_corr} shows that these corrections shift the metallicities the most in the MSTO region, while the right panel shows that applying the AD correction narrows the spread in mono-metallic tracks along the full CMD. We also plot the [M/H] distribution of the RGB stars vs. the MSTO (see \autoref{fig:adc_hist}). To aid more direct comparison, we also plot the [M/H] of the RGB with "MSTO-like" errors (meaning for each RGB star we randomly sample a MSTO star error then resample the given RGB metallicity from a Gaussian with a sigma equivalent to the new error. Because RGB metallicities are more well-measured, this demonstrates the overall spread in RGB metallicities if they had the same uncertainties as fainter MSTO stars). Before the correction, it is clear that the MSTO stars are shifted to lower metallicities by $\sim 0.2$ dex while after the correction the distributions are significantly more consistent, as expected.

Using atomic diffusion corrected metallicities allows us to group stars across the CMD by their birth metallicity, not their observed atmospheric metallicity.  This enables us to more accurately group stars with similar abundances and better understand the processes that enrich each population. We include both the raw and atomic diffusion corrected values for the metallicities in our catalog and we use the corrected values in all analyses forthcoming. 

\subsection{Perspective rotation}\label{sec:persective rotation}

\begin{figure}[t]
\plotone{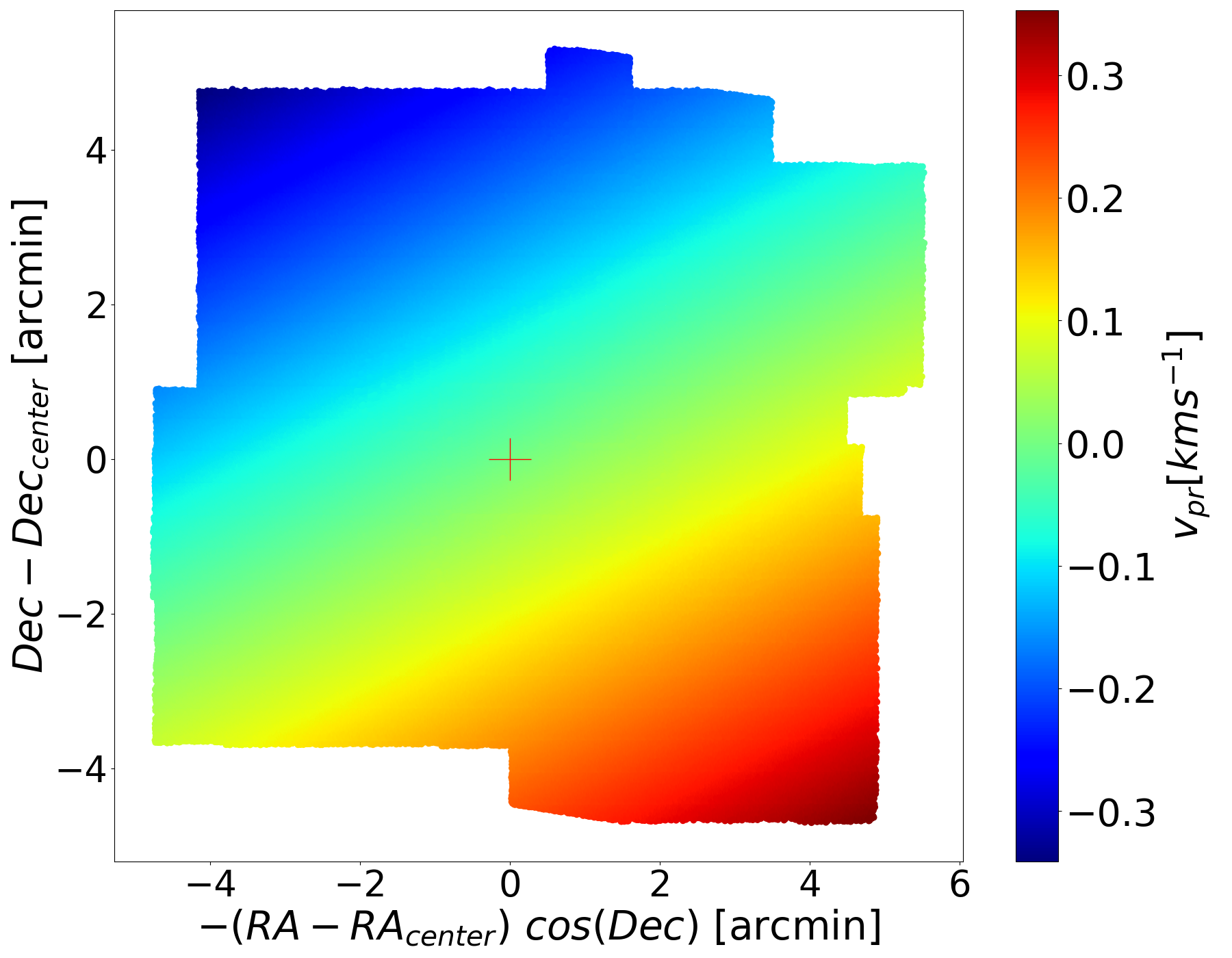}
\caption{\textbf{Perspective rotation.} The figure shows the position of the data color coded with the perspective rotation value for each star. The red cross is indicating the center and the velocity values increase from the bottom left to the top right corner of the plot.\label{fig:vpr}}
\end{figure}

$\omega$~Cen takes up a large angle on the plane of the sky. Therefore, there is a non-negligible apparent rotation caused by the different projection of the space motion at different positions on the sky. This is known as perspective rotation \citep{VanDeVen_2006}. To correct for it we need to subtract from the line-of-sight velocity the perspective rotation term:
\begin{eqnarray}\label{eq: vpr}
    v_{pr} [\mathrm{km~s^{-1}}] = 1.3790\cdot 10^{-3} \cdot D \cdot \nonumber \\\left( -\Delta RA\cdot cos(Dec) \cdot\mu_{RA}^{sys} + \Delta Dec\cdot \mu_{Dec}^{sys}\right) 
\end{eqnarray}
with the distance to the center $\Delta RA$, $\Delta Dec$ in units of arcmin, distance to the cluster D = (5.43 $\pm$ 0.05)~kpc \citep{Baumgardt2021}, proper motion in RA $\mu_{RA}^{sys} =  (3.25 \pm 0.022)$~mas~$\mathrm{yr^{-1}}$ and proper motion in Dec $\mu_{Dec}^{sys} =  ( -6.746 \pm 0.022)$~mas~$\mathrm{yr^{-1}}$ from \citet{Vasiliev2021}.

The resulting velocity calculated with \autoref{eq: vpr} is shown in \autoref{fig:vpr}, and the maximum value for the perspective rotation in our dataset is \SI{0.35}{\kilo\meter\per\second}.

\subsection{Catalog}\label{sec:catalog}

\begin{figure}[b]
\plotone{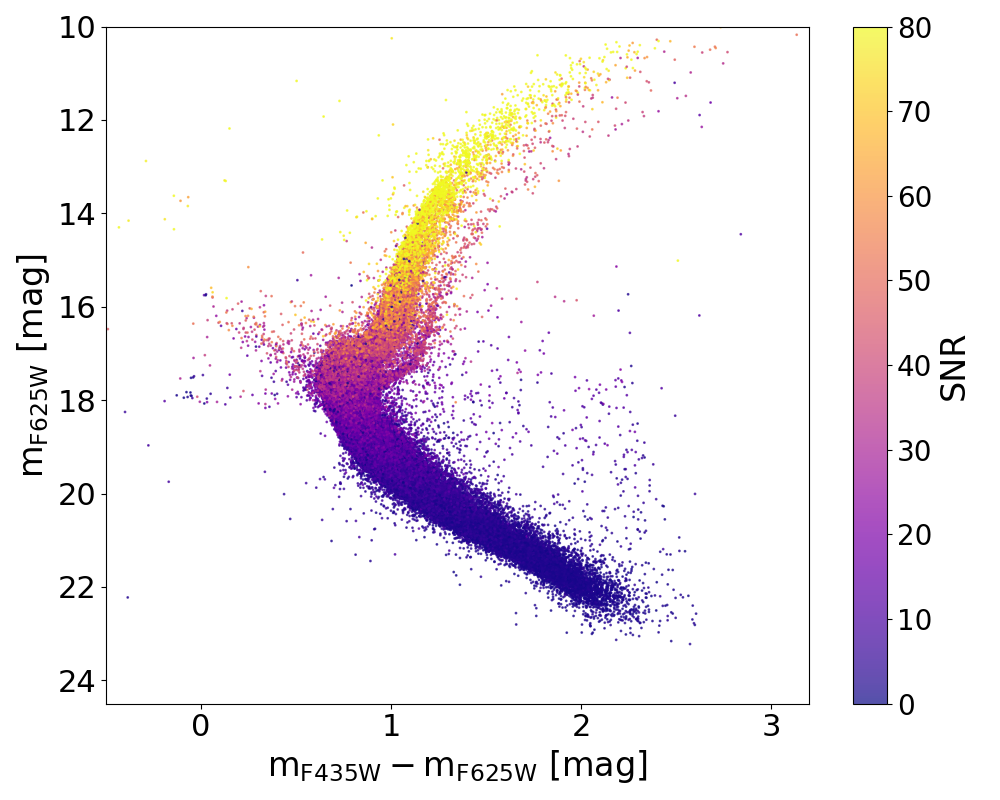}
\caption{\textbf{SNR in the catalog}. The CMD of the stars in the catalog color coded with their SNR as calculated from \textsc{spexxy}. \label{fig:cmd Snr}}
\end{figure} 
\begin{figure*}[t]
\plottwo{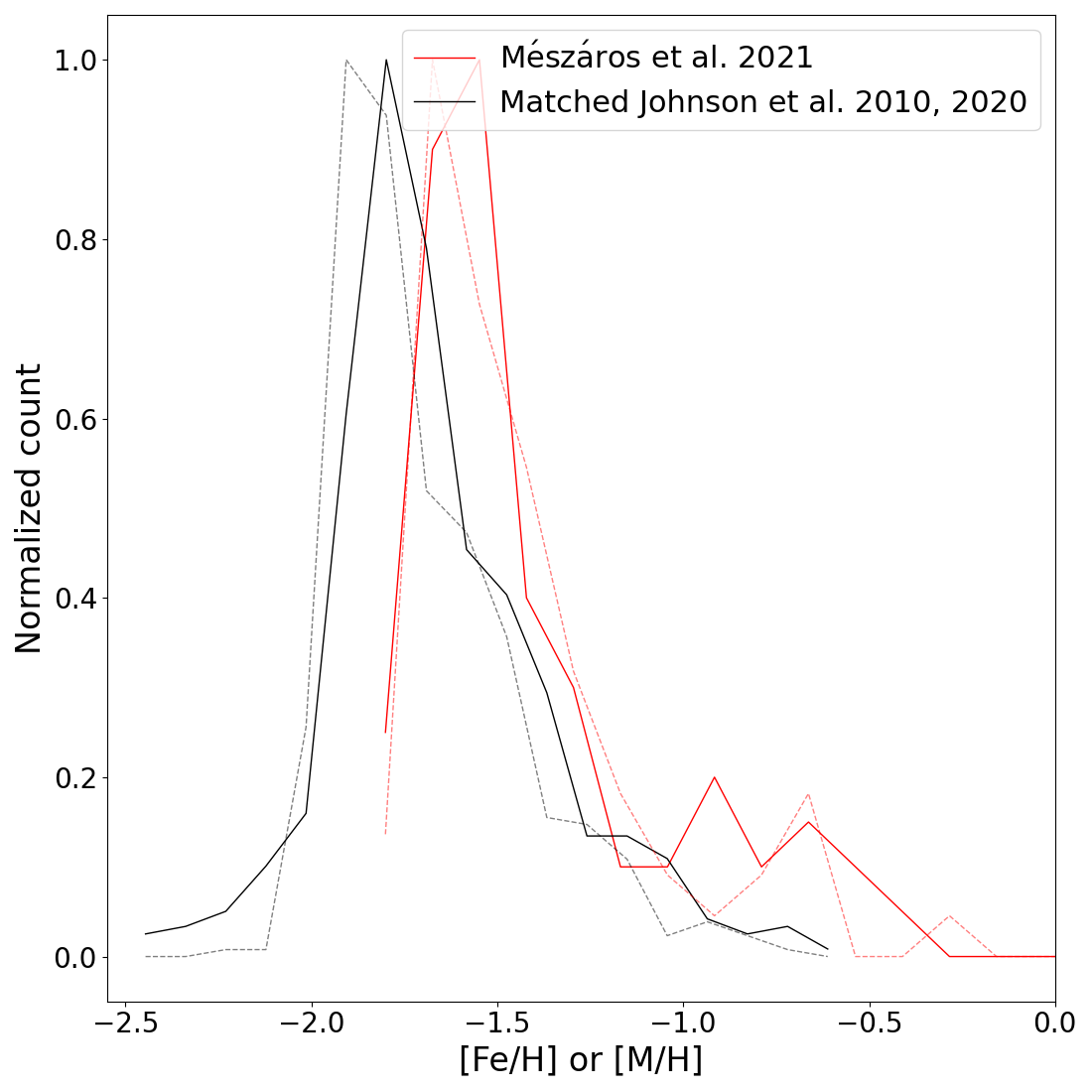}{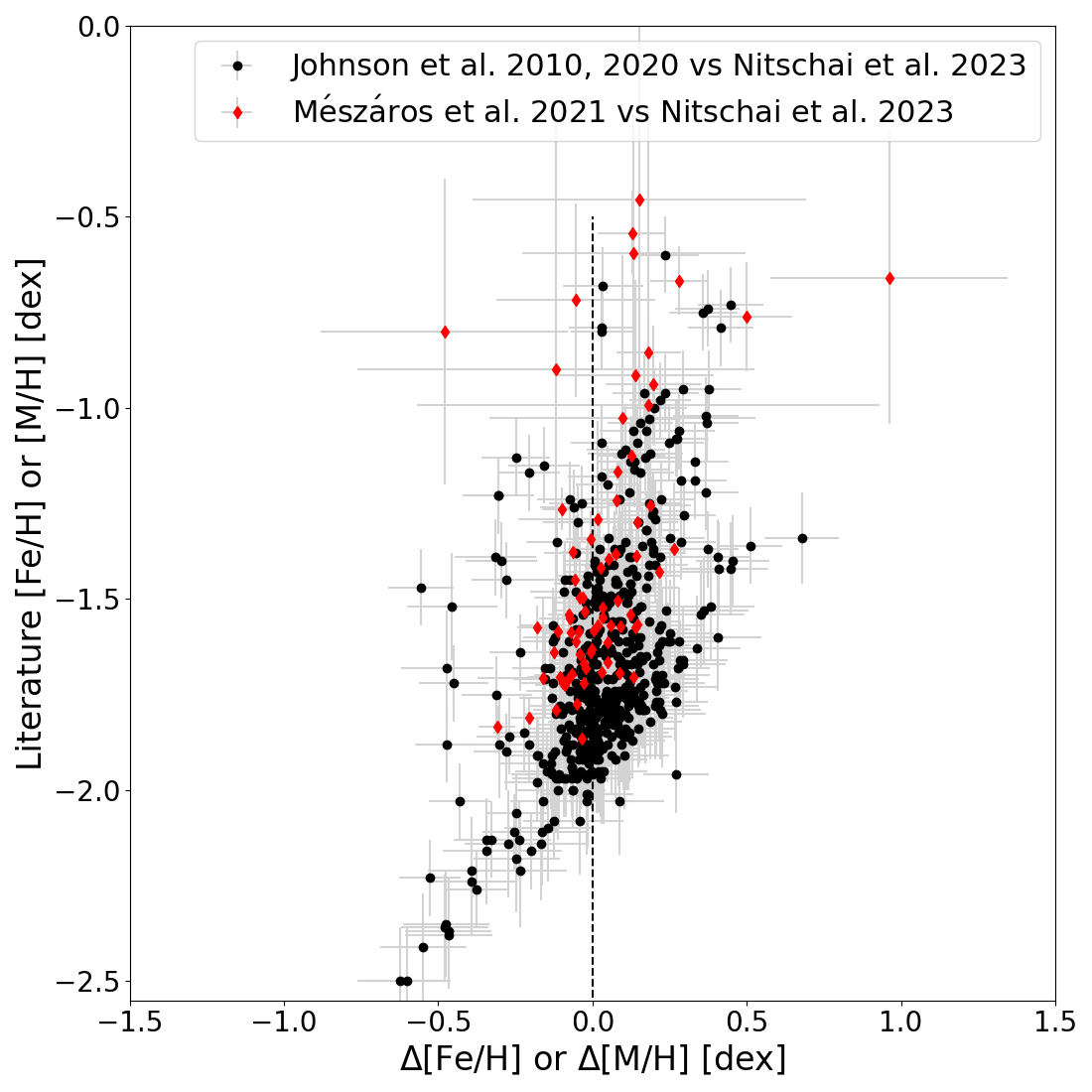}
\caption{\textbf{Comparison to literature values}. On the left is the [Fe/H] distribution for the \citet{Johnson_2010, Johnson_2020}, the black solid line, and from the MUSE results, the black dashed line, for all the matched stars in both catalogs. In red is the [M/H] distribution for the matched stars of the \citet{Meszaros_2021} data, solid line, and the MUSE data. On the right, we plot the literature values against the difference between the two [Fe/H] or [M/H] values. The black dashed line indicates where both values agree. \label{fig:Feh}}
\end{figure*}

Having performed all the data reduction, stellar spectra extraction, and analysis steps described in Section~\ref{sec:data}, \ref{sec:methods} and \ref{sec:results}, we have a final spectroscopic catalog with 342,797 stars (GTO: 58,143, GO: 272,633 and combined GO/GTO: 12,021). All the parameters given in this catalog are described in \autoref{sec:columns} and the SNR of the stars in the CMD is shown in \autoref{fig:cmd Snr}. The catalog will be published together with this paper in a machine-readable format.

Since 1,432 GTO stars are not in the \citet{Anderson2010} catalog, they do not have magnitudes in the F625W and F435W filters, and therefore no atomic diffusion correction (ADC) could be performed. We will update the ADC for these stars in a subsequent paper using a new \textit{HST} catalog (M. H\"aberle et al. in prep) that includes new data acquired under the program GO-16777 (PI: A. Seth).

Our catalog has no quality cuts, as each science case can decide the best cuts for its own purpose. However, the quality cuts we suggest are as follows:
\begin{itemize}
    \item Membership probability: $P_{M} > $ 95 per cent
    \item Magnitude accuracy during extraction with \textsc{PampelMuse} $\geq 0.6$
    \item Reliability: $ R \geq 0.5$
    \item Distance to MUSE IFU edge:  $\geq$ 5 pixel (1\arcsec)
    \item Signal-to-Noise ratio: SNR $> 10$ (see \autoref{sec:PampelMuse test} and \autoref{sec:snr test}).
    \end{itemize}

The above cuts yield 156,871 stars and of these, only 143 GTO stars lack photometry in the \citet{Anderson2010} catalog. These 156,871 stars have a mean metallicity error of 0.15~dex, a median $\rm v_{los}$ error of \SI{6.26}{\kilo\meter\per\second} and a median $\rm T_{eff}$ error of \SI{102}{\kelvin}. In more detail, the mean metallicity error for MS stars (\SI{18}{\mag} $\rm < mag_{F625W}<$\SI{22}{\mag}) is 0.174~dex and for RGB stars (\SI{16}{\mag} $<\rm  mag_{F625W}<$\SI{10}{\mag}) 0.031~dex. We decided on these cuts in order to have reliable results but still keep as many stars as possible, stricter cuts will give smaller uncertainties. The SNR cut of 10 is consistent with all our tests in \autoref{sec:PampelMuse test}, \autoref{sec:logg} and \autoref{sec:snr test}, which show that below that value biases because of different assumptions and setups can play a role. The distance to the edge is chosen to verify that the stars are not outside or exactly at the edge of the IFU where only part of the starlight can be extracted. The other parameters are reasonable cuts for the extraction accuracy, reliability, and membership, but stricter cuts can be applied for higher precision.

Additionally, the median systemic LOS velocity, corrected for the perspective rotation and sign the above quality cuts, is $\rm v_{los} = (232.99 \pm 0.06)~km~s^{-1}$ with the error found using bootstrapping and even without the reliability quality cut the result does not change. This value is close to previous values, e.g.
$\rm (232.7\pm 0.2)~km~s^{-1}$ \citet{Baumgardt2018} using ESO/VLT, Keck spectra, and literature values, being just slightly out of the 1$\sigma$ range, while $\rm (234.28 \pm 0.24)~km~s^{-1}$ \citet{Baumgardt2019} where they used \textit{Gaia} DR2 data, which due to crowding do not have many stars in the central region where we observe, being outside of the 3$\sigma$ range due to the small errors.

One should note that these are quality cuts based on our spectroscopic analysis, while the photometry and astrometry are added from the \textit{HST} catalogs \citep{Sarajedini2007, Anderson2010}. We are preparing a new astrometric and photometric catalog (M. H\"aberle et al. in prep) that will provide updated values for the magnitudes and positions.

\subsection{Literature Comparison}\label{sec:comparison}

In this subsection, we present a first comparison of our findings with previous works. For the comparison, we use other previous studies that use spectroscopic data and not just photometry.

To compare our [M/H] values from \textsc{spexxy} (without AD correction) to literature values for [Fe/H] from RGB stars \citep[e.g.:][]{Johnson_2010, Johnson_2020} we need to transform [M/H] to [Fe/H] values. We use the following formula \citep{Salaris_1993}:
\begin{equation}
    [Fe/H] = [M/H] - \log(0.638 \cdot 10^{[\alpha/Fe]} +0.362)
\end{equation}
with $[\alpha/Fe]$ set to 0.3 in \textsc{spexxy}. This transformation accounts for the contribution of the assumed $\alpha$ enhancement to our assumed [M/H] values, while the literature values derive separate abundances for several $\alpha$ elements \citep{Johnson_2010}.  

We match the \citet{Johnson_2010, Johnson_2020} catalogs to ours using only the brightest stars (F625W $<$ 14 mag) by finding the closest star that is no more than \SI{1}{\arcsec} away, in order to find real matches but not have too strict criteria and eliminate matches due to slight offsets in the coordinates. The [Fe/H] of the 524 matched stars are shown in \autoref{fig:Feh}. The measurements are strongly correlated with a mean offset between them of 0.03 dex and a scatter of 0.17 dex, with the literature values being slightly more metal-rich but still close to zero showing no strong systematic offset and the scatter representing the statistical discrepancy. We note that our matching criteria and differences in astrometry could result in some mismatched stars.  Unfortunately, this cannot easily be improved using photometry, as there are no common filters between the catalogs.  In addition, we do not find as low metallicity stars as in \citet{Johnson_2020}, which seem to have the biggest deviation from our data.

Further, we also match our [M/H] values to the metallicity from \citet{Meszaros_2021} where they used APOGEE measurements. Since, most of the spectroscopic surveys, like \textit{Gaia} and APOGEE, have problems with crowding in the central cluster region where we observe, there is almost no overlap. However, we could match 74 stars, and the results are also shown in \autoref{fig:Feh}. The measurements have a mean offset between them of 0.06 dex and a scatter of 0.28 dex.

\section{Conclusion}\label{sec: concl}

In this work, we present an extensive catalog of stars obtained with 87 new MUSE pointings plus the already existing 10 WFM and 6 NFM GTO pointings for $\omega$~Cen, covering out to the half-light radius. 

We describe the data and the analysis steps used to create the final catalog of stars. First, we reduce the data and extract individual spectra of stars using \textsc{PampelMuse} \citep{Kamann2013}. Afterwards, we use the spectral fitting routine \textsc{spexxy} to measure the LOS velocities, effective temperatures, and metallicities of these stars. We perform multiple tests including exploiting a large number of repeated measurements to verify our results. We provide several parameters to quantify the reliability of the results and also include several necessary corrections to our measurements. We end up with 342,797 stars after all of our analysis steps without any quality cuts. However, for most use cases we suggest several quality cuts: SNR $>$ 10, membership probability $\rm P_{M} >$ 0.95 \%, Magnitude accuracy $\geq$ 0.6\,mag, distance to the IFU edge $\geq$ 5 and Reliability $R \geq$ 0.5. Thus, 156,871 stars meet all these criteria.

Finally, we do a first analysis of the metallicity distribution comparing it to previous works and find that they are consistent with our findings and calculate the median systematic LOS to be $\rm v_{los} = (232.99 \pm 0.06)~km~s^{-1}$. A more detailed analysis of the metallicities as well as the LOS velocity information will be presented in a separate subsequent paper. In addition, a new \textit{HST} catalog with proper motions photometry for millions of stars covering most of the MUSE data is in preparation (M. H\"aberle et al. in prep.). Combining both, the astrometric and spectroscopic catalog for $\omega$~Cen, will allow us to measure the ages of the stars and identify the subpopulations of the cluster using photometry and metallicity information (C. Clontz et al. in prep.). Moreover, using the spectroscopic data we can also measure individual abundances by combining spectra of the same population and directly measuring the absorption lines of different elements (S. Di Stefano et al. in prep.), which will further constrain the multiple populations and their formation scenario. Combining the information on the stellar populations and kinematics of a large number of stars in $\omega$~Cen will provide a clear picture of its formation history. 

\section*{Acknowledgments}
The work is based on observations collected at the European Southern Observatory under ESO program 105.20CG.001. We thank the ESO staff for their excellent support. Also based on observations made with the NASA/ESA Hubble Space Telescope, obtained from the data archive at the Space Telescope Science Institute. STScI is operated by the Association of Universities for Research in Astronomy, Inc. under NASA contract NAS 5-26555. This research made use of NASA’s Astrophysics Data System. AS, CC, and MAC acknowledge support from \textit{HST} grant GO-16777. We acknowledge funding from the Deutsche Forschungsgemeinschaft (grant LA
4383/4-1 and DR 281/35-1) and from the German Ministry for Education and Science (BMBF Verbundforschung) through grants 05A14MGA, 05A17MGA, and 05A20MGA. SK acknowledges funding from UKRI in the form of a Future Leaders Fellowship (grant no. MR/T022868/1). We thank the anonymous referee for the helpful and constructive comments.

\facilities{VLT:Yepun (MUSE), \textit{HST} (ACS)}

\software{\textsc{PampelMuse} \citep{Kamann2013}, \textsc{spexxy} v5.2.21 \url{https://github.com/thusser/spexxy}, \textsc{Astropy} v5.2.1
\citep{astropy:2013, astropy:2018, astropy:2022}, \textsc{clumpy} \citep{Kimmig_2015}, \textsc{Matplotlib} v3.7.1 \citep{Hunter:2007, Caswel2023}, \textsc{Pandas} v1.5.3 \citep{mckinney-proc-scipy-2010, reback2020pandas}, \textsc{Numpy} v1.20.3 \citep{harris2020array}, \textsc{SciPy} v.1.10.1 \citep{2020SciPy-NMeth}, MUSE pipeline v2.8.3 \citep{MUSE_pipeline, Weilbacher2020}
}
\appendix

In the Appendix, we include further information and elaborate on checks we performed on our data. 

\section{Data Conditions}\label{sec:cond data}
Since our data were taken over an extended period of longer than a year, the atmospheric conditions were also varying.
In \autoref{tab:conditions} we summarize the conditions and observing date of the individual GO pointings used in this analysis. We requested an airmass less than 1.4 and seeing below $0\farcs8$, which were almost always fulfilled.

\startlongtable
\begin{deluxetable}{ccccc}
\tablecaption{Observing Conditions \label{tab:conditions}}
\tablewidth{0pt}
\tablehead{
\colhead{OB} &  \colhead{Date}& \colhead{Average Airmass} & \colhead{Average Seeing}}
\startdata
1.1  & 17 February 2021 & 1.263 & 0.897 \\
1.2 & 17 February 2021 & 1.246 & 0.777\\
1.3 & 17 February 2021 & 1.204 & 0.733\\
1.3\footnote{OB 1.3 was observed twice since it had a wrong offset causing a gap in our dataset.} & 2 September 2022 & 2.144 & 0.577\\
2.1  & 16 March 2021 & 1.102 & 1.203 \\
2.2 & 16 March 2021 & 1.092 & 1.133\\
2.3 & 16 March 2021 & 1.087 & 1.357 \\
3.1  & 16 March 2021 & 1.088 & 1.097 \\
3.2 & 16 March 2021 & 1.093 & 1.127 \\
3.3 & 16 March 2021 & 1.103 & 1.360 \\
4.1  & 8 April 2021 & 1.096 & 0.433\\
4.2 & 8 April 2021 & 1.089 & 0.500\\
4.3 & 8 April 2021 & 1.086 & 0.470 \\
5.1  & 14 March 2021 & 1.088 & 1.140 \\
5.2 & 14 March 2021 & 1.094 & 1.110\\
5.3 & 14 March 2021 & 1.103 & 1.133 \\
5.4 & 14 March 2021 & 1.116 & 1.273 \\
6.1  & 8 April 2021 & 1.088 & 0.503\\
6.2 & 8 April 2021 & 1.093 & 0.527\\
6.3 & 8 April 2021 & 1.101 & 0.497 \\
6.4 & 8 April 2021 & 1.113 & 0.547 \\
7.1  & 18 May 2021 & 1.209 & 0.567 \\
7.2 & 18 May 2021 & 1.180 & 0.547 \\
7.3 & 18 May 2021 & 1.155 & 0.553\\
7.4 & 18 May 2021 & 1.135 & 0.450\\
8.1  & 18 May 2021 & 1.089 & 0.59\\
8.2 & 18 May 2021 & 1.086 & 0.573\\
8.3 & 18 May 2021 & 1.087 & 0.687\\
9.1  & 30 June 2021 & 1.088 & 0.883 \\
9.2 & 30 June-1 July 2021 & 1.093 & 0.907\\
9.3 & 1 July 2021 & 1.101 & 1.147\\
9.4 & 1 July 2021 & 1.113 & 1.163\\
10.1  & 4 July 2021 & 1.106 & 0.937\\
10.2 & 4 July 2021 & 1.119 & 0.877 \\
10.3 & 4 July 2021 & 1.119 & 0.843\\
10.4 & 4 July 2021 & 1.157 & 0.840\\
11.1 & 31 March 2022 & 1.295 & 0.62\\
11.2 & 31 March 2022 & 1.253 & 0.540 \\
11.3 & 31 March 2022 & 1.217 & 0.50 \\
11.4 & 31 March 2022 & 1.186 & 0.4340 \\
12.1  & 9 February 2022 & 1.325 & 0.557\\
12.2 & 9 February 2022 & 1.279 & 0.677\\
12.3 & 9 February 2022 & 1.239 & 0.637\\
12.4 & 9 February 2022 & 1.206 & 0.700\\
13.1 & 9 February 2022 & 1.171 & 0.507\\
13.2 & 9 February 2022 & 1.148 & 0.703 \\
13.3 & 9 February 2022 & 1.128 & 0.563\\
13.4 & 9 February 2022 & 1.113 & 0.693\\
14.1 & 9 February 2022 & 1.099 & 0.723 \\
14.2 & 9 February 2022 & 1.091 & 1.020\\
14.3 & 9 February 2022 & 1.087 & 0.650\\
14.4 & 9 February 2022 & 1.085 & 0.780\\
15.1 & 10 February 2022 & 1.250 & 0.653\\
15.2 & 10 February 2022 & 1.215 & 0.727\\
15.3 & 10 February 2022 & 1.184 & 0.823\\
15.4 & 10 February 2022 & 1.159 & 0.597\\
16.1 & 10 February 2022 & 1.133 & 0.740\\
16.2 & 10 February 2022 & 1.117 & 0.600\\
16.3 & 10 February 2022 & 1.103 & 0.433\\
16.4 & 10 February 2022 & 1.094 & 0.897\\
17.1 & 27 February 2022 & 1.358 & 0.350\\
17.2 & 27 February 2022 & 1.308 & 0.307\\
17.3 & 27 February 2022 & 1.265 & 0.293\\
17.4 & 27 February 2022 & 1.228 & 0.250\\
18.1 & 27 February 2022 & 1.090 & 0.387\\
18.2 & 27 February 2022 & 1.086 & 0.413\\
18.3 & 27 February 2022 & 1.086 & 0.357\\
18.4 & 27 February 2022 & 1.089 & 0.520\\
19.1 & 27 February 2022 & 1.100 & 0.437 \\
19.2 & 27-28 February 2022 & 1.147 & 0.463\\
19.3 & 28 February 2022 & 1.201 & 0.500\\
19.4 & 28 February 2022 & 1.173 & 0.627\\
20.1 & 28 February 2022 & 1.143 & 0.493 \\
20.2 & 28 February 2022 & 1.125 & 0.397 \\
20.3 & 28 February 2022 & 1.110 & 0.380\\
20.4 & 28 February 2022 & 1.100 & 0.380 \\
21.1 & 28 February 2022 & 1.360 & 0.413\\
21.2 & 28 February 2022 & 1.087 & 0.453 \\
21.3 & 28 February 2022 & 1.087 & 0.373\\
21.4 & 28 February 2022 & 1.091 & 0.307\\
22.1 & 2 March 2022 & 1.379 & 0.640\\
22.2 & 2 March 2022 & 1.326 & 0.517\\
22.3 & 2 March 2022 & 1.280 & 0.567\\
22.4 & 2 March 2022 & 1.241 & 0.877\\
23.1 & 2 March 2022 & 1.114 & 0.487\\
23.2 & 2 March 2022 & 1.102 & 0.493\\
23.3 & 2 March 2022 & 1.094 & 0.437\\
23.4 & 2 March 2022 & 1.089 & 0.463\\
\enddata
\tablecomments{The observing conditions for all GO observations used in our work. OB: the name/numbering of the individual pointings, Date: the date of observation, average airmass: the mean airmass at the start of each of the three rotated exposures, and average seeing: the mean seeing at the start of the three rotated exposures for one OB. OB 1.3 was observed twice since the offset of the pointing was not correct and created a gap in our continuous coverage.}
\end{deluxetable}

\section{\textsc{PampelMuse} tests}\label{sec:PampelMuse test}

We verify that slightly changing the \textsc{PampelMuse} setup, does not influence our results for stars with  SNR$> $ 10. For these tests, we use two example pointings where a different setting might have been more appropriate. First, we use a round PSF, fixing the ellipticity to 0, for OB 15.3, since the free Moffat PSF we generally use seems unnecessarily complex as it is round anyway. For the second  example, we use OB 21.2 and masked certain layers in the wavelength range that appear to have something wrong with them, bad pixels from the data reduction, since their values drop significantly compared to the rest of the spectrum. The results show no significant difference in the \textsc{spexxy} output for an SNR $>$ 10, even though the new setup should be better for these two cases. Therefore, we keep the setup consistent for all OBs as described in \autoref{sec:PampelMuse}.

\section{Surface gravity log(g)}\label{sec:logg}

In this section of the Appendix we investigate the bias we have due to the fixed log(g) parameter from one single isochrone for most stars (see \autoref{sec:spexxy}).
\begin{figure*}[t]
\gridline{\fig{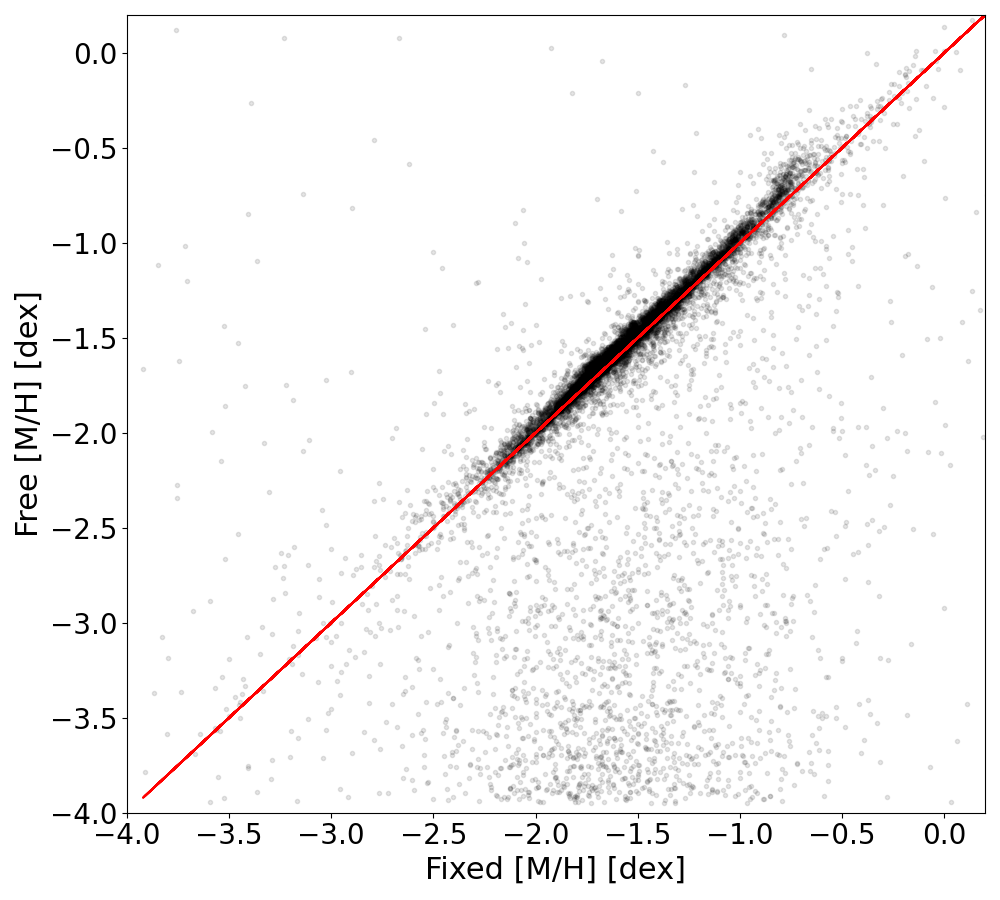}{0.4\textwidth}{(a) [M/H] comparison}}
\gridline{\fig{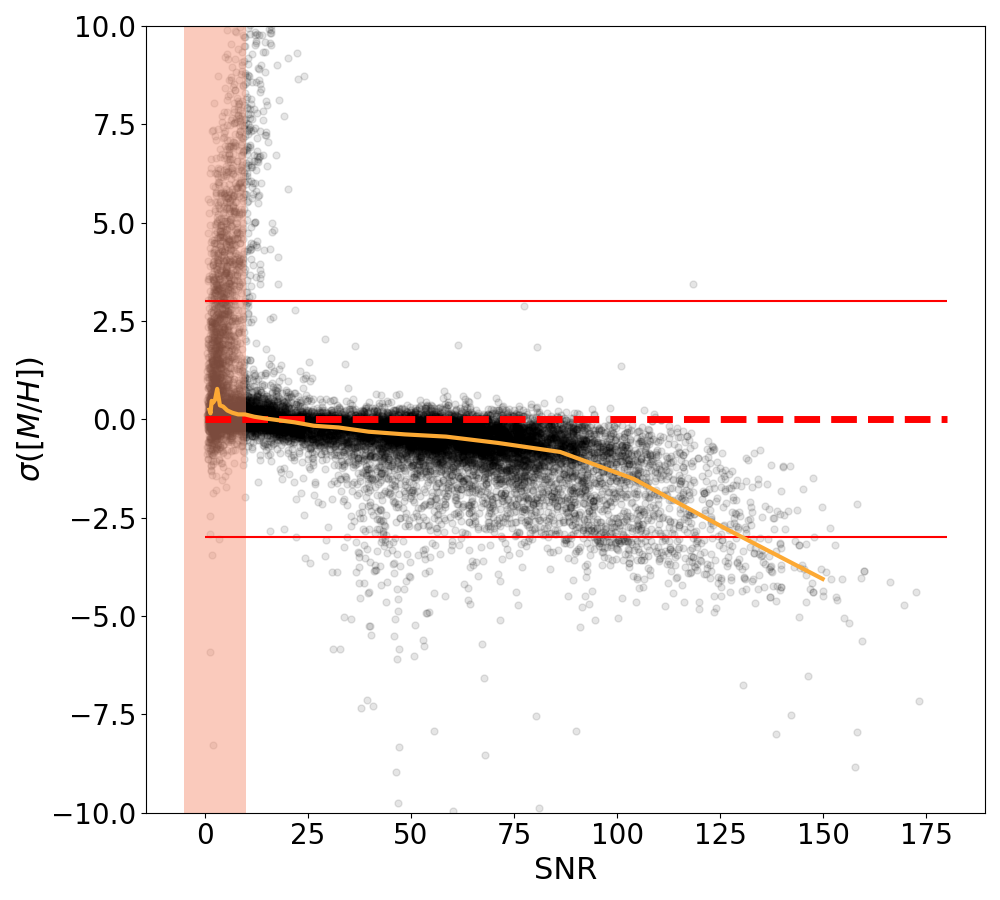}{0.4\textwidth}{(b) [M/H] sigma discrepancy}
          \fig{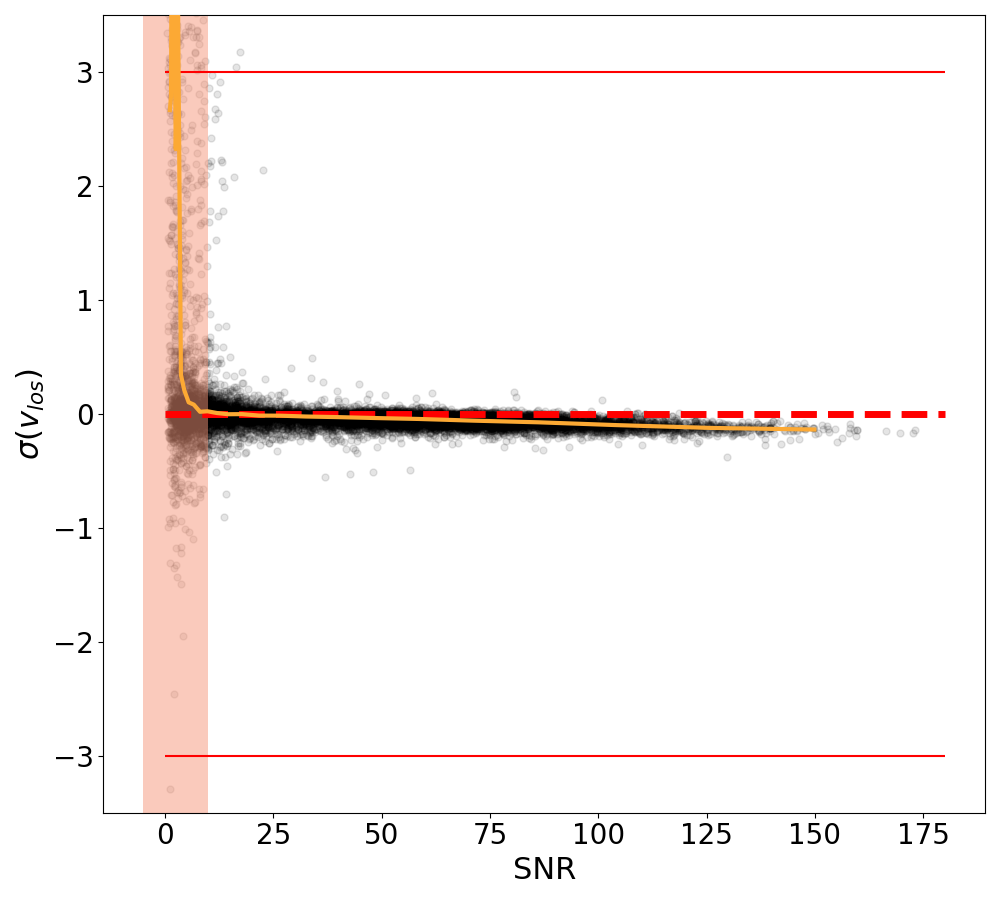}{0.4\textwidth}{(c) $\rm v_{los}$ sigma discrepancy}}
\gridline{\fig{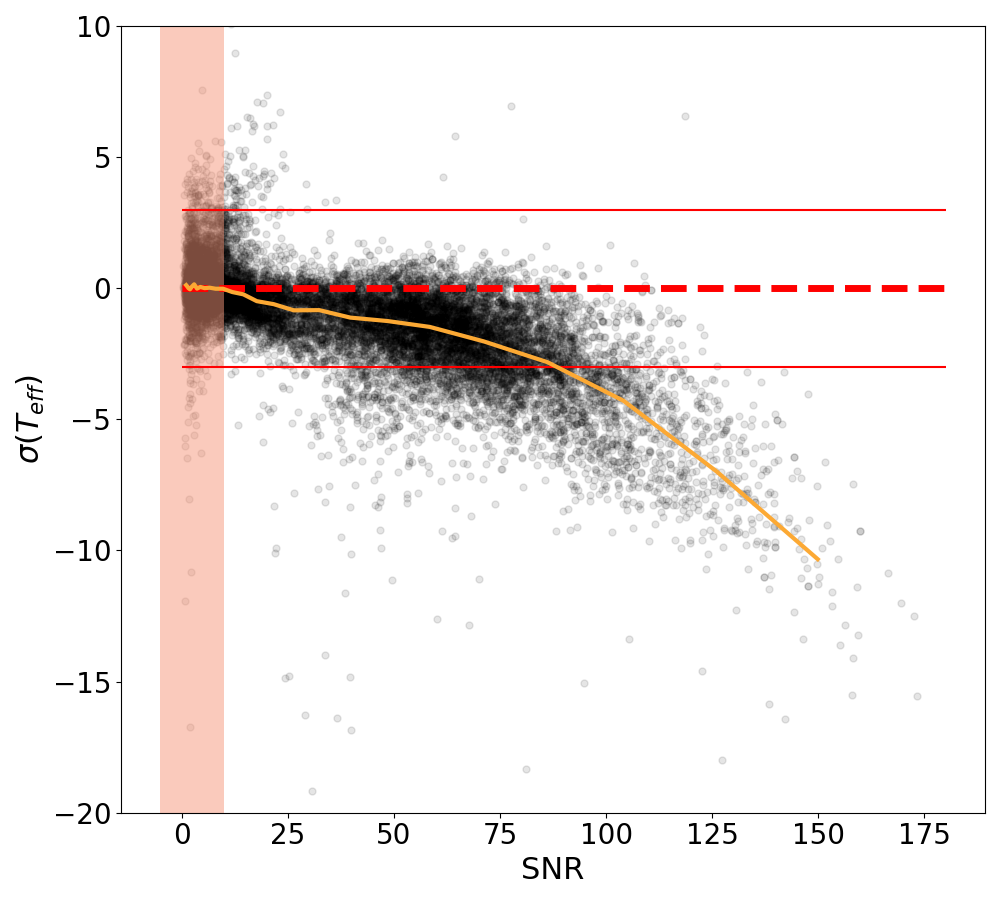}{0.4\textwidth}{(d) $\rm T_{eff}$ sigma discrepancy}
          \fig{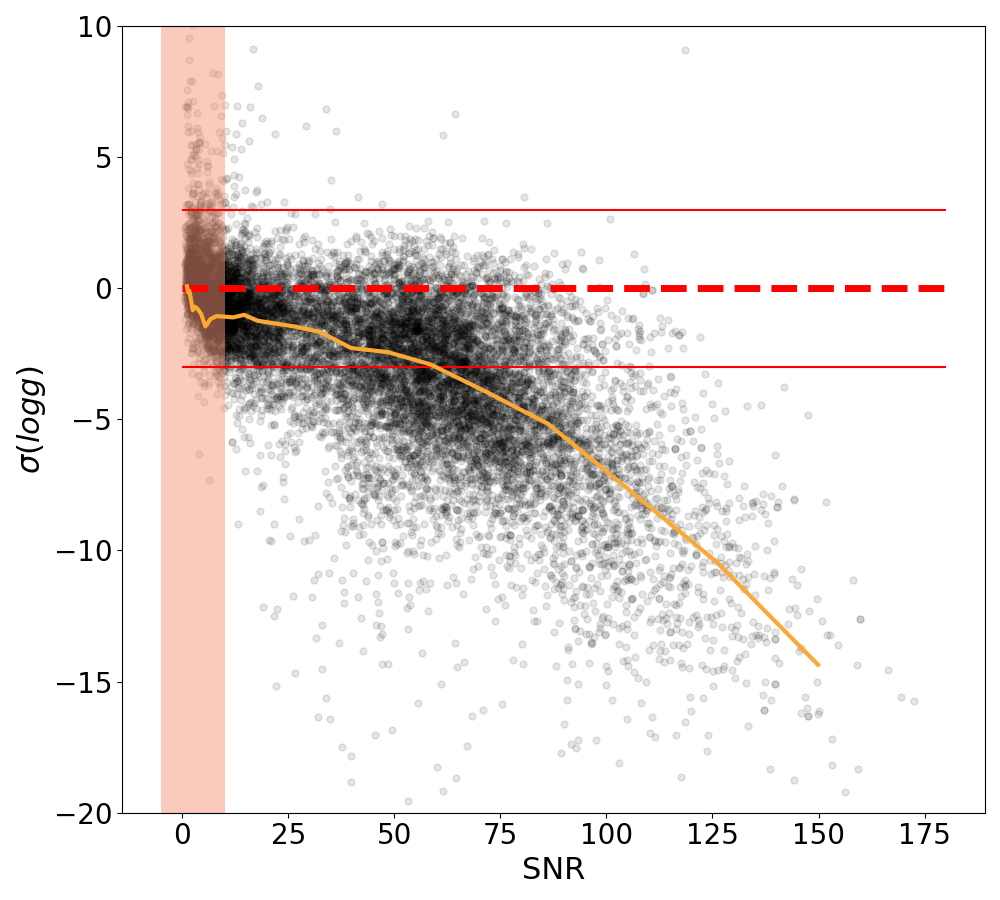}{0.4\textwidth}{(e) log(g) sigma discrepancy}}

\caption{\textbf{Discrepancy between spectral parameters with free and fixed log(g)} (a) The one-to-one comparison between the [M/H] values, red line the 1 to 1 ratio, (b)-(e); The sigma deviations for [M/H], $\rm v_{los}$, $\rm T_{eff}$ and log(g). We divide the difference between the two results with the scale error in our catalog, except for the log(g) where we take the \textsc{spexxy} error for the free log(g) run. The dashed red line is at 0, the solid red lines at the 3 $\sigma$ limits, and the orange solid line is the median of all stars, while the red shaded area is the SNR limit of 10.\label{fig:free logg}}
\end{figure*}

\subsection{Free log(g)}\label{sec:logg free}

First, we investigate how allowing log(g) to vary freely for all stars during the fit with \textsc{spexxy}, would change our results. For that, we look at a subset of our data including RGB and MS stars, which gives us a complete representation of the magnitude and SNR range.

The deviations between the free and fixed log(g) fit are shown in \autoref{fig:free logg}. Most of the results are within the 3~$\sigma$ range for a SNR $>$ 10. The deviations at high SNR are because of the smaller errors at that level. We can see that all parameters tend to get lower when the log(g) values are left  free, but not as significant for [M/H] and $\rm v_{los}$ where they almost all stay within the 3~$\sigma$ range. This also proves that the log(g) parameter does not affect the metallicity calculation strongly.

\subsection{Metal rich stars}\label{sec:logg metal rich}

\begin{figure}[h!]
\plottwo{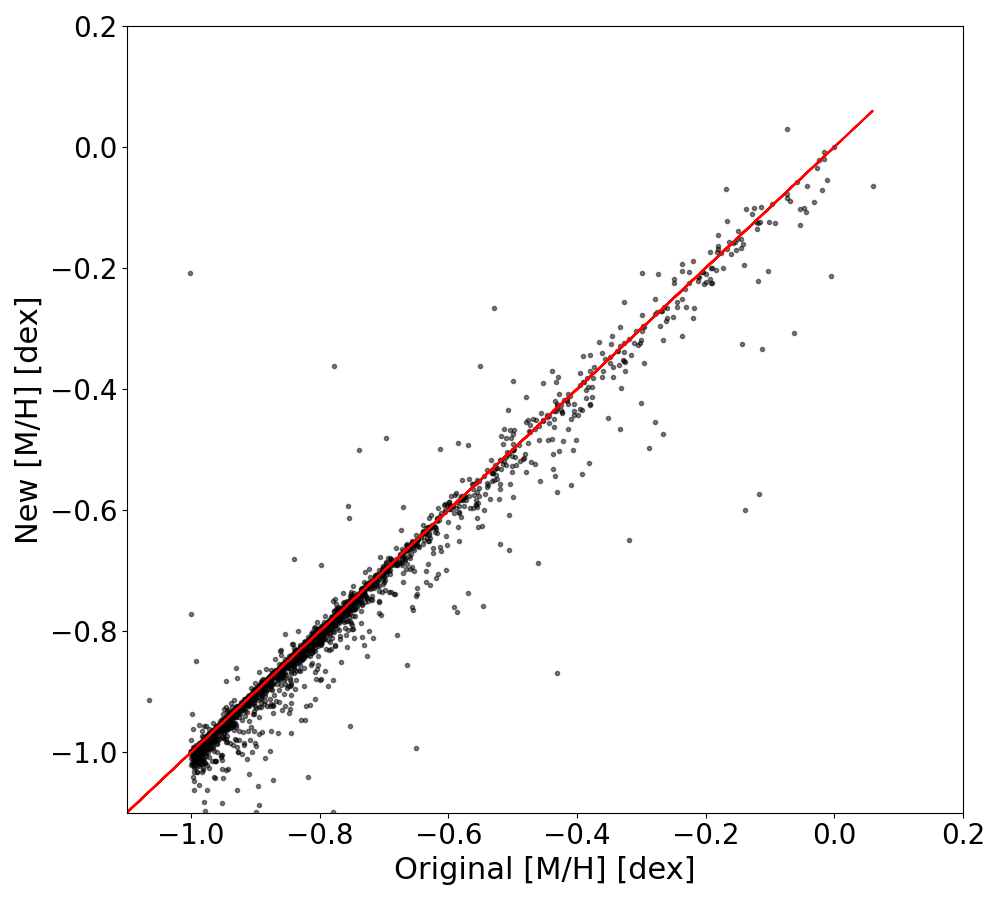}{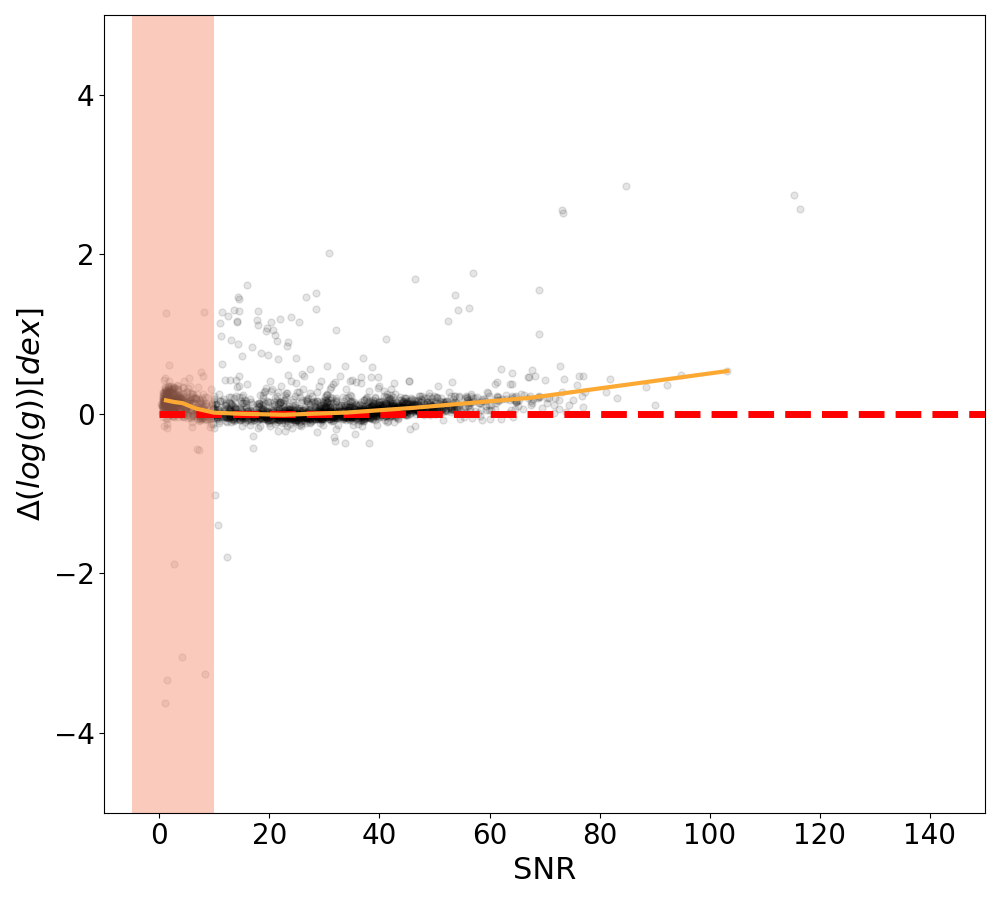}
\caption{\textbf{Metal rich}. On the left plot, the one-to-one comparison of the [M/H] is shown, with the red solid line being the 1 to 1 ratio. On the right is the difference in log(g) between the two isochrones, with the dashed line indicating no difference at all, the orange line the median difference of our data, and the red shaded area the SNR limit of 10. \label{fig:metal rich logg}}
\end{figure}

Since the cluster is known to have multiple populations, we know that one isochrone for finding the log(g) is not ideal. To investigate the bias we have because of that, we test one extreme population, the metal rich stars with [M/H]$>-1$~dex, and use a different isochrone, more representative for them, and assign them new log(g) values.

In \autoref{fig:metal rich logg} the [M/H] results are shown and the difference between the log(g). In general, there seems to be just a small difference in log(g) even with a different isochrone and a tight one-to-one correlation between the [M/H], reassuring us that the exact isochrone used for the initial guesses and log(g) values is not that significant.

\subsection{Literature log(g)}\label{sec:logg lit}

\begin{figure}[t]
\epsscale{0.5}
\plotone{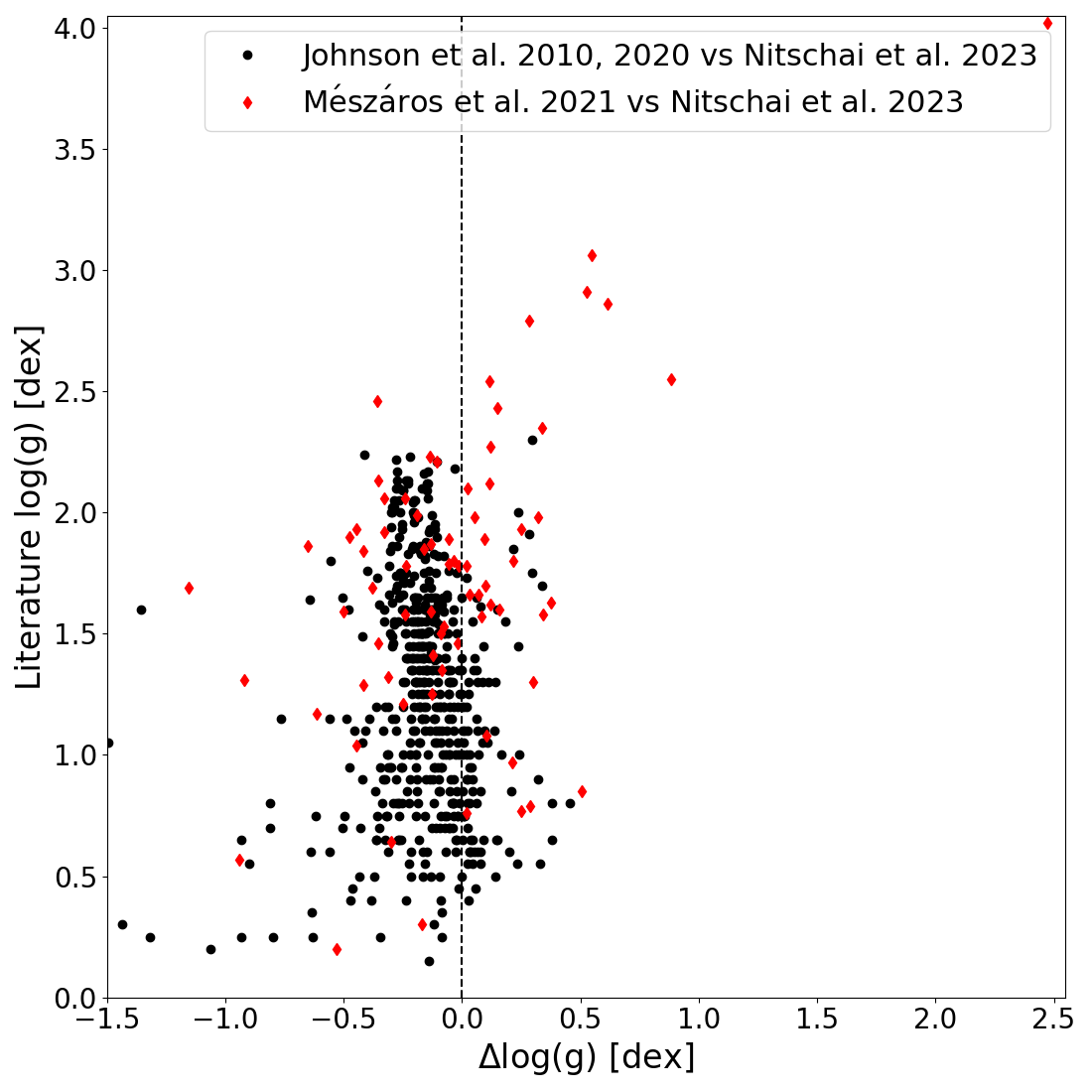}
\caption{\textbf{Literature comparison log(g).} We plot the Literature values against the difference between Literature and our log(g) values.\label{fig:lit logg}}
\end{figure}
Using the matching between \citet{Johnson_2010, Johnson_2020, Meszaros_2021} and our catalog, described in \autoref{sec:comparison}, we also compare our surface gravity values with the literature. Almost all values are for a fixed log(g) from the isochrone described in \autoref{sec:spexxy}.

The log(g) of the matched stars are shown in \autoref{fig:lit logg}. The measurements have a mean offset between the \citet{Johnson_2010, Johnson_2020} of 0.18 dex and a scatter of 0.26 dex, with the literature values having slightly lower log(g) values. With the \citet{Meszaros_2021} values the mean offset is 0.04 dex and a scatter of 0.46 dex. This shows that even though our initial isochrone is not a perfect match for all stars, because of multiple populations and binaries, it still gives reasonable log(g) values since they are comparable to previous studies. In addition, the metallicity values also are within the expected literature range (see \autoref{sec:comparison}) even though log(g) is fixed, hence this is another indication that there is no strong effect, as already suggested in \citet{Husser_2016, Kamann_2016} and in \autoref{sec:logg free}, \autoref{sec:logg metal rich}.

\section{SNR test}\label{sec:snr test}

To make sure that we have no bias in our metallicity and velocity measurements for stars with different SNR, we performed another test, adding different noise values to the spectra and redoing the \textsc{spexxy} fit.

\begin{figure}[h]
\plottwo{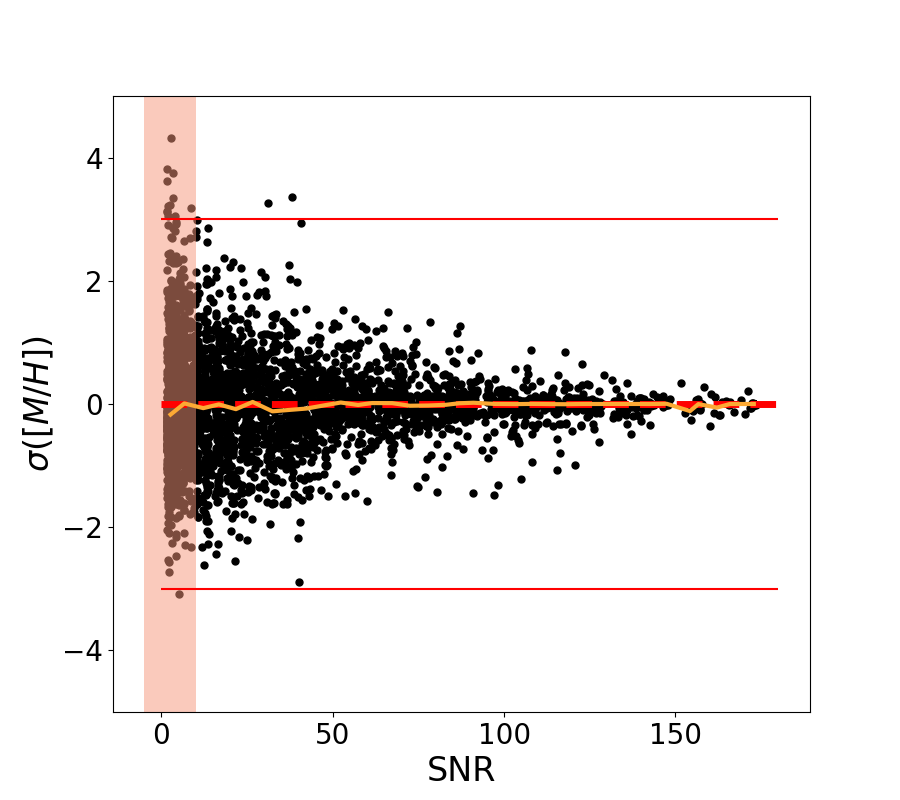}{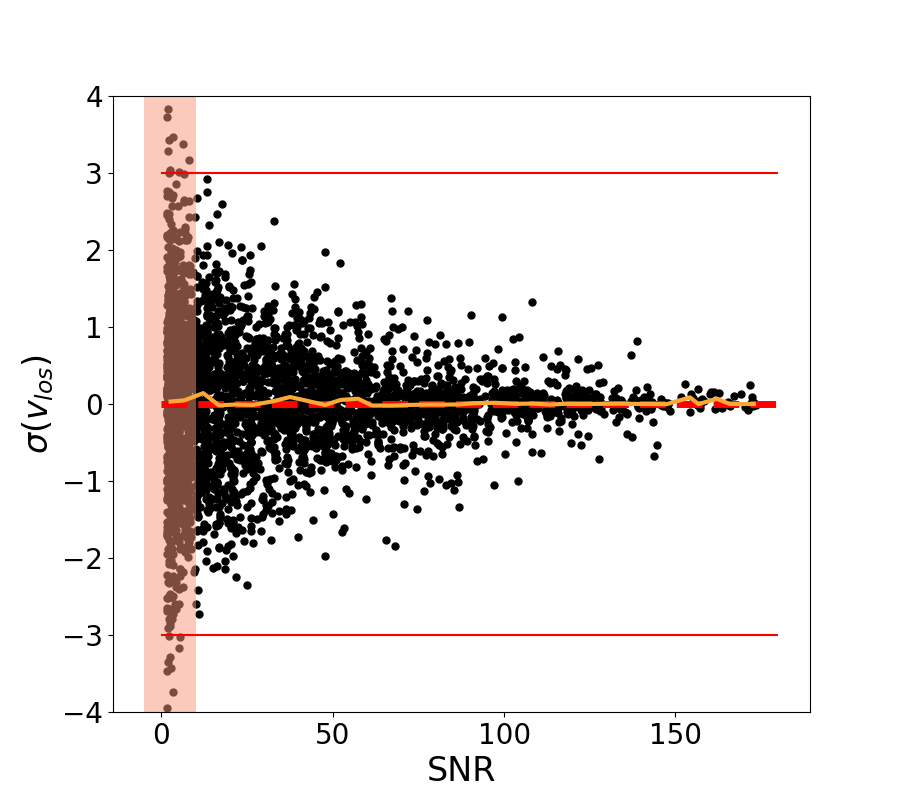}
\caption{\textbf{SNR test}. The left plot shows on the y-axis the $\sigma$ difference between the original result and the different [M/H] results after adding noise to the spectra, and the x-axis is the new SNR with the added noise. Similar for the right plot but for the $\rm v_{los}$ values. The light red shaded areas indicate SNR below 10, the red solid horizontal lines are the 3$\sigma$ difference, the red dashed horizontal lines are at 0, and the orange solid lines are the median values for our measurements as a function of SNR. \label{im:test snr}}
\end{figure}

We selected randomly 147 spectra with SNR above 28 (being not on the HB) and added Gaussian noise, creating 3,999 spectra with SNR between 1 and the original SNR. The difference between the original \textsc{spexxy} results and the new ones with more noise increases as expected for lower SNR, and the output errors also increase. The $\sigma$ difference as a function of the new SNR is shown in \autoref{im:test snr}. Almost all results are within the 3$\sigma$ range for SNR $>$ 10 and we do not see any bias towards higher or lower values for lower SNR.

The $\sigma$ difference is calculated using the error directly from \textsc{spexxy}, if we also apply the error correction as described in \autoref{sec:errors}, the difference even decreases.

\section{Catalog Columns}\label{sec:columns}

\autoref{tab:columns} lists all parameters given in the spectroscopic catalog described in this paper. The entries in the catalog are sorted by increasing radius from the center.

\startlongtable
\begin{deluxetable}{l|cccc}
\tablecaption{Catalog Columns \label{tab:columns}}
\tablewidth{0pt}
\tablehead{
\colhead{Column Name} &  \colhead{Description}& \colhead{More Details} & \colhead{Units}}
\startdata
MUSE & Identifier of stars in spectroscopic catalog & & \\
RVel & Line-of-sight velocity ($\rm v_{los}$) & \textsc{spexxy} output & \SI{}{\kilo\meter\per\second} \\
errRVel & \textsc{spexxy} Line-of-sight velocity uncertainty ($\epsilon_{v_{los}}$)& \textsc{spexxy} output & \SI{}{\kilo\meter\per\second}\\
sRVel & Scaling factor for $\epsilon_{v_{los}}$ & \autoref{sec:errors} & \\
e\_RVel & Scaled error $\rm v_{los}$ & \autoref{sec:errors} & \SI{}{\kilo\meter\per\second}\\
prv & Perspective rotation &  see \autoref{sec:persective rotation} and \citet{VanDeVen_2006} & \SI{}{\kilo\meter\per\second}\\
prvRVel & $\rm v_{los}$ corrected for perspective rotation & see \autoref{sec:persective rotation} and \citet{VanDeVen_2006} & \SI{}{\kilo\meter\per\second}\\
Teff & Effective Temperature ($\rm T_{eff}$) & \textsc{spexxy} output & K\\
errTeff & \textsc{spexxy} Effective Temperature uncertainty ($\epsilon_{T_{eff}}$)& \textsc{spexxy} output & K\\
sTeff & Scaling factor for $\epsilon_{T_{eff}}$ &\autoref{sec:errors} & \\
e\_Teff & Scaled error $\rm T_{eff}$ & \autoref{sec:errors}& K\\
loggfix & Fixed surface gravity ($\rm log(g)$) & Isochrone output (\autoref{sec:spexxy}) & dex\\
loggfree & Fitted surface gravity ($\rm log(g)$) & \textsc{spexxy} output & dex\\
MH & Metallicity ([M/H]) & \textsc{spexxy} output & dex\\
MHadc & Metallicity corrected for AD & see \autoref{sec:atomic_diffusion} & dex\\
errMH & \textsc{spexxy} Metallicity uncertainty ($\epsilon_{[M/H]}$) & \textsc{spexxy} output & dex\\
sMH & Scaling factor for $\epsilon_{[M/H]}$ & \autoref{sec:errors} & \\
e\_MH & Scaled error [M/H] & \autoref{sec:errors} & dex\\
SNR & Signal-to-noise ratio (SNR) & \textsc{spexxy} output & \\
ffit & Fixed or free log(g) in \textsc{spexxy} & 0 free and 1 fixed & \\
edge & Distance of star to OB edge & \textsc{PampelMuse} output & pixel\\
Magacc & Relative accuracy of recovered magnitude & \textsc{PampelMuse} output & \\
& from spectrum extraction &&\\
SpecFlag & Spectrum extraction quality flag & \textsc{PampelMuse} output & \\
Rel & Reliability Parameter $R$& \autoref{sec:reliability} & \\
Num & Number of times the star was observed & \autoref{sec:multi} & \\
Data & Observing program (GO/GTO) the star belongs to & GTO, GO or both (GO\_GTO) & \\
Rad & Radius from cluster center & $\sqrt{\left(( RA -RA_{c})cos(Dec)\right)^2 + \left( Dec-Dec_{c}\right)^2}$ & deg\\
RAdeg & Right Ascension & \citet{Sarajedini2007, Anderson2008},& deg\\
&&\citet{Anderson2010}&\\
DEdeg & Declination & \citet{Sarajedini2007, Anderson2008},& deg\\
&&\cite{Anderson2010}&\\
435mag & Magnitude in the F435W filter (Vega) & \citet{Anderson2010} & mag\\
625mag & Magnitude in the F625W filter (Vega) & \citet{Anderson2010} & mag\\
435magC& F435W-band photometry (Vega) & $ \rm A_{V}$ and $ \rm R_{V}$  Correction Applied & mag\\
625magC & F625W-band photometry (Vega) & $ \rm A_{V}$ and $ \rm R_{V}$ Correction Applied & mag\\
e\_435mag & RMS scatter of single-exposure F435W observation & \citet{Anderson2010} & mag\\
e\_625mag &  RMS scatter of single-exposure F625W observation & \citet{Anderson2010} & mag\\
NBf & Number of F435W images where star was found & \citet{Anderson2010} & \\
NRf &  Number of F625W images where star was found & \citet{Anderson2010} & \\
probRV & Membership probability & radius and velocity (\autoref{sec:members}) & \\
probRVM & Membership probability & radius, velocity and [M/H]& \\
Flag & Quality Flag & 1=True, 0=False, see \autoref{sec:catalog} & \\
HST10Flag & \textit{HST} quality flag for \citet{Anderson2010} & 1=reliable, 0=unreliable  photometry/astrometry & \\
\enddata
\tablecomments{$\mathrm RA_c$ = 13:26:47.24 and $\mathrm Dec_c$ =-47:28:46.45 \citep{Anderson2010}. For the parameters edge, Magacc, SpecFlag and Rel we provide the minimum (maximum for SpecFlag) value when combining multiple measures, see \autoref{sec:multi}.}
\end{deluxetable}

\bibliography{paper}{}
\bibliographystyle{aasjournal}

\end{document}